\documentclass[conference]{IEEEtran}
\IEEEoverridecommandlockouts
\usepackage{cite}
\usepackage{amsmath,amssymb,amsfonts}
\usepackage{subcaption}
\usepackage{graphicx}
\usepackage{caption}
\usepackage{multirow}
\usepackage[Procedure]{algorithm}
\usepackage[noend]{algpseudocode}
\usepackage{graphicx}
\usepackage{textcomp}
\usepackage{xcolor}
\def\BibTeX{{\rm B\kern-.05em{\sc i\kern-.025em b}\kern-.08em
    T\kern-.1667em\lower.7ex\hbox{E}\kern-.125emX}}

\definecolor{orange}{rgb}{1,0.5,0}

\newcommand{\ac}[1]{{#1}}

\usepackage{ulem}  % strike through  \sout{ }
\usepackage{url}

\usepackage{booktabs}

\begin{document}

\title{Discriminating Equivalent Algorithms via Relative Performance\\
\thanks{Financial support from the Deutsche Forschungsgemeinschaft (German Research Foundation) through grants GSC 111 and IRTG 2379 is gratefully acknowledged.}
}

\author{\IEEEauthorblockN{Aravind Sankaran}
\IEEEauthorblockA{\textit{IRTG - Modern Inverse Problems} \\
\textit{RWTH Aachen University}\\
Aachen, Germany \\
aravind.sankaran@rwth-aachen.de}
\and
\IEEEauthorblockN{Paolo Bientinesi}
\IEEEauthorblockA{\textit{Department of Computer Science} \\
\textit{Ume\r{a} Universitet}\\
Ume\r{a}, Sweden \\
pauldj@cs.umu.se}

%\author{\IEEEauthorblockN{Author 1}
%	\IEEEauthorblockA{\textit{Institute} \\
%		Address \\
%		Contact Information}
%	\and
%	\IEEEauthorblockN{Author 2}
%	\IEEEauthorblockA{\textit{Institute} \\
%		Address \\
%		Contact Information}

%\and
%\and
%\and
}

\maketitle
\thispagestyle{plain}
\pagestyle{plain}

\begin{abstract}

In scientific computing, it is common that a mathematical expression can be computed by many different algorithms
(sometimes over hundreds), each identifying a specific sequence of library calls.
%\p{I don't like this opening sentence. The fact that one target computation can be implemented in many different ways is
%not limited to scientific computing. Also, one computation is not ``evaluated''; that does not mean anything. You need
%to find a suitable wording, and then use it consistently. One target ``problem'' can be ``solved'' by many different
%``algorithms''. Or one target ``computation'' can be ``implemented'' in many different ``ways'' (or ``programs''?). Or
%target ``calculation''?, ``computed''?, ``algorithms''? ...}
Although mathematically equivalent, those algorithms might exhibit significant differences in terms of performance. However in practice, due to fluctuations, there is not one algorithm that consistently performs noticeably better than the rest.
%\p{This second sentence can be made stronger. As it is, it sounds rather obvious. I would say that out of these many
%  algorithms, also due to fluctuations/jitter/noise, it is common that there is not one of them that consistently
%  performs noticeably better than the rest. On the contrary, several algorithms exhibit comparable performance, i.e.,
%  their distributions (-explain-) have considerable overlap.}
For this
reason, with this work we aim to identify not the one best algorithm, but
the subset of algorithms that are reliably faster than the rest. To this
end, instead of using the usual approach of quantifying the
performance of an algorithm in absolute terms, we present a
measurement-based clustering approach to sort the algorithms into equivalence (or performance) classes using pair-wise comparisons. 
 We show that this approach, based on relative performance, leads to robust
 identification of the fastest algorithms even under noisy system conditions. Furthermore,
 it enables the development of practical machine learning models for automatic algorithm selection. 
\end{abstract}


\section{Introduction}

%\ar{
\ac{Given a set $\mathcal{A}$ of mathematically equivalent algorithms (i.e., algorithms that in exact arithmetic would
  all return the same output),}
%\p{maybe we should specify ``i.e., in exact arithmetic they would all return the exact same output,'' ?}
we aim to identify the subset
$\mathcal{F} \subseteq \mathcal{A}$ that contains all those algorithms that are ``equivalently'' fast to one another, and
``noticeably'' faster than the algorithms in the subset $\mathcal{A} \setminus \mathcal{F}$. We will clarify the meaning of ``equivalent'' and
``noticeable'' shortly; for now, we simply state that in order to identify $\mathcal{F}$, 
we develop a measurement-based approach that assigns a higher score to the algorithms in
$\mathcal{F}$ compared to those in $\mathcal{A} \setminus \mathcal{F}$.  
%The subset $\mathcal{F}$ is specific to a given computing architecture, operating system, and run time settings. 
In order to capture the performance of an algorithm, we compute a relative score that compares the current algorithm with respect to the fastest algorithm(s) in $\mathcal{A}$. We refer to such scores as ``relative performance estimates''.

\begin{figure}[h!]
	\centering
		\begin{subfigure}[b]{0.5\textwidth}
		\includegraphics[width=1\linewidth]{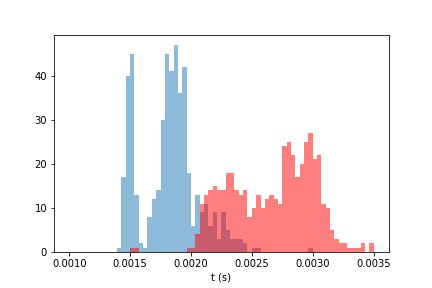}
		\caption{Algorithms having noticeable difference in distributions.
                 % \p{Somewhere you should point out that how separated two distributions must be to be
                 %   considered ``noticeably different'' is a user-defined parameter. For instance, one could decide that
                 %   the distributions have to be entirely disjoint, or that they should not overlap for more than 20\%
                 %   of ...} \as{Is it important to mention about this here? I can explain about this in the experiment
                 %   section.}\p{notice that I say ``somewhere'', as opposed to ``here'' ;-)}
                }
		\label{fig:diff} 
	\end{subfigure}

	\begin{subfigure}[b]{0.5\textwidth}
		\includegraphics[width=1\linewidth]{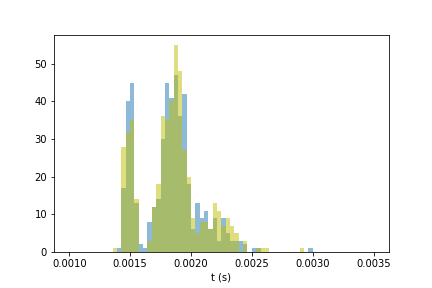}
		\caption{Algorithms having significant overlap of distributions.}
		\label{fig:eq}
	\end{subfigure}
	
	\caption{Five hundred measurements of three solution algorithms (Yellow, Blue, Red) for the Ordinary Least Square problem.}
	\label{fig:1}
\end{figure}

It is well known that execution times are influenced by many factors, and that repeated measurements, even with same
input data and \ac{compute environment},
%\p{what do we mean by ``same cache conditions''?}
often result in different execution times~\cite{peise2014cache,hoefler2010characterizing,peise2012performance}.
%\p{This is the place to introduce the concept of distribution.}
\ac{Therefore, comparing the performance of any two algorithms involves comparing two sets of measurements (or ``distributions'')}. In common practice, \ac{time distributions}
%\p{and/or ``distributions'' (once we introduce the concept)}
are summarized into \ac{few} statistical estimates (such as minimum or median execution
time, \ac{possibly in combination with standard deviations or quantiles}),
%\p{should we say ``into FEW statistical estimates (such as min or median ..., possibly in combination with std dev or
%  quartiles(?) ...)'' ?}
which are then used to compare and rank algorithms~\cite{peise2019elaps,hoefler2010characterizing,parasec}
%\p{if you can find additional references (besides this one), I am sure the reviewers would appreciate}
However, when system noise begins to have a significant impact on program execution,  the \ac{common statistical quantities}
%\p{``a single quantity''?  also possible ``FEW numbers/quantities''?}
cannot reliably capture the profile of the time distribution~\cite{hoefler2015scientific};
%\p{``time distribution'' has not been introduced yet.}
\ac{as a consequence, when time measurements are repeated, the ranking of algorithms would most likely change at every repetition and this makes the development of reliable machine learning models for automatic algorithm comparisons difficult.} 
%\p{instead of ``it is not possible'', let's rephrase as ``as a consequence, if/when time measurements are repeated,
%  THIS FACT/THAT FACT is/are extremely likely to happen. In other words, the ranking of algorithms would change at every
%  repetition ... / (Or:) In other words, the ranking of algorithms would be very sensitive to small changes in the
%  observations (measurements)''}
The lack of consistency in ranking stems from not considering the \ac{possibility that two algorithms can be equivalent} when comparing their
performance. 
%\p{you mean ``not considering the possibility that two algorithms are equivalent'' ?}
In order for one algorithm to be faster (or slower) than the other, there should be ``noticeable'' difference in their
time distributions; \ac{for instance, comparing the blue and red time distributions in Fig.~\ref{fig:diff}, the blue algorithm could be considered reliably faster than the red one}.
On the other hand, the performance of algorithms is comparable if their distributions are ``equivalent'' or have significant overlap (Figure \ref{fig:eq}). 
Therefore, the comparison of two algorithms will yield one of the three outcomes: \ac{faster, slower, or equivalent.}
%\p{Given that we're comparing time distributions, I'd replace ``better'' with ``faster'' (or ``noticeably
 % faster'', or better yet ''reliably faster'', ``consistently faster''), ``slower'' (``reliably slower'', ``consistently
 % slower''), and ``equivalent''.}
In this paper, we \ac{describe a methodology} to use this three-way comparison to sort the set of algorithms $\mathcal{A}$ into performance classes by merging the ranks of algorithms whose distributions are significantly overlapping with one another.
%  and to construct a ranking,  that is not only consistent (robust) with noisy system conditions, but also allows for equivalent algorithms to be assigned to the same rank, which is important when one wishes to optimize over additional performance criteria.
%

The algorithms in $\mathcal{A}$ represent different, alternative ways of computing the same mathematical expression.
%\p{notice my comment in the abstract. Here: ``alternative ways to compute an expression'' -- this is good, if used
%  consistently. In Fig.1: ``solution algorithms to .. problem''; also good, but different. In the opening sentence of
%  the abstract: ``computation evaluated by algorithms''; this I don't like.}
%In exact arithmetic, those algorithms would all return the same quantity.
%\p{I suggested to move this sentence to the opening of the intro}
For instance,  consider the following expression which appears in an image restoration application\cite{tirer2018image}: $y := H^{T}y
+ (I_n - H^{T}H)x$;  where $H$ is a square matrix in $\mathbb{R}^{n \times n}$ , $x$ and $y$ are vectors in $\mathbb{R}^{n}$, and $I_n$ is an identity matrix. 
If the product $H^TH$ is computed explicitly, the code would perform a $\mathcal{O}(n^3)$ matrix-matrix multiplication; by contrast, by applying
distributivity\footnote{In general, distributivity does not always lead to lower FLOP count.}, one can rewrite this
assignment as $y := H^{T}(y - Hx) + x$, obtaining an alternate algorithm which computes the same expression by
using only matrix-vector multiplications, for a cost of $\mathcal{O}(n^2)$. In this example, the two algorithms differ
in the order of magnitude of floating point operations (FLOPs), hence noticeable differences in terms of execution times are naturally expected. 
However, \ac{with increasing parallelism or efficient memory access patterns, it is possible that an algorithm with higher FLOP counts result in faster
executions~\cite{barthels2019linnea}.}
%\p{This sentence seems to suggest that it is due to the parallelism the fact that more FLOPS might result in faster
%  executions, but this is not the case. The difference has to do with the efficiency of those flops. (E.g.: flops in BLAS 2 vs BLAS 3 routines}
%However, two algorithms may significantly differ in execution times even if they perform the same number of FLOPs, and it is even possible that higher FLOP counts result in faster executions~\cite{barthels2019linnea}.

%As mentioned earlier, to determine which of two algorithms is the fastest, their distributions of execution times should be compared.  The result will depend on the specific sizes of input matrices (operand sizes). Figure [1] shows the distributions of the two algorithms for image restoration for two different dimensions of $H$. When $H$ is small, both the algorithms are equally fast and we want to assign them both to $\mathcal{F}$. If the dimension of $H$ is large, the gains of distributivity starts showing and now only the second algorithm should be in $\mathcal{F}$. 

In practice, one has to consider and compare more than just two algorithms.  For instance, for the Generalized Least Square problem $y := (X^TS^{-1}X)^{-1}X^{T}S^{-1}z$, where $X \in \mathbb{R}^{n \times m}$, $S \in \mathbb{R}^{n \times n}$ and $z \in \mathbb{R}^{n}$, it is possible to find more than 100 different algorithms that compute the same solution and that differ not more than 1.4x in terms of FLOP count.\footnote{Julia code for 100 equivalent algorithms of the Generalized Least Square problem is available here github.com/as641651/Relative-Performance/tree/master/Julia-Code/GLS/Julia/generated}  
%the product $H^{T}H$ could be computed both by the kernel $gemm$, which implements matrix matrix multiplications, and by calling a different kernel $syrk$, which instead computes a symmetric rank-k update.
All these different algorithms arise by changes in the sequence of BLAS calls due to properties of input matrices (such as symmetric positive definite, lower or upper triangular, etc.), different parenthesizations for matrix chains, identifying common sub-expressions, etc. \cite{psarras2019linear}. The number of equivalent algorithms increases exponentially when considering the possible splits of computations among different resources (such as CPU and GPUs). In this work, we consider linear algebra expressions which occur at the heart of many scientific computing applications. We consider solution algorithms generated by the Linnea framework~\cite{barthels2019linnea}, which accepts linear algebra expressions as input, and generates a family of algorithms (in the form of sequential Julia code~\cite{julia}) consisting of (mostly, but not limited to) sequences of BLAS and LAPACK calls. 

\ac{In this paper, we develop a measurement-based methodology for reliable discrimination (or clustering) of algorithms. This methodology can also be extended for use in the (more challenging) scenario in which it is not feasible to time all algorithms. In that case, in order to get closer in identifying the set of fastest algorithms $\mathcal{F}$ within a given time budget (for instance, 6 hours), one should make as many ``intelligent'' measurements as possible. To this end, the clustering is done on a sample of algorithms and further measurements are sampled in such a way that the probability of being in the set $\mathcal{F}$ is higher. The details of this intelligent sampling, which can be achieved by using our methodology, is beyond the scope of this article. }

\textbf{\textit{Contributions:} }The efficient computation of mathematical expressions
%\p{Here you use ``computation of expressions'', as you used earlier on. Maybe then stick to this?}
is critical both for complex
simulations and for time-critical computations done on resource-constrained hardware \cite{towardsEdgeComputing}
(e.g., instantaneous object detection for autonomous driving applications \cite{connectedvehicles}). A framework that
identifies the best algorithm should take into consideration at least the systematic noise factors, such as  regular
interrupts, impacts of  other applications that are simultaneously sharing the resources, etc. Moreover, we observed that
multiple algorithms can yield similar performance profiles. Therefore,  we develop a methodology for robust identification of not one, but a set of  fast algorithms for a given operational setting. From the resulting set of fast algorithms, one may seek to select an algorithm based on additional performance metrics such as energy~\cite{paiseSankaran}. To this end, we use the bootstrapping technique~\cite{bootstrap} to rank the set of equivalent algorithms $\mathcal{A}$ into clusters, where more than one algorithm can obtain the same rank. The robust rankings produced by our methodology can also be used as ground truth to train realistic performance models that can predict the ranks without executing all the algorithms\footnote{The modeling and prediction of relative performance is the objective of our future work, and is out of the scope of this article.}.

\textit{\textbf{Organization}}: In Sec.~\ref{sec:rel}, we highlight the challenges and survey the state of art. In
Secs.~\ref{sec:torel} and~\ref{sec:met}, we introduce the idea behind relative performance, and describe a methodology
to obtain relative scores, respectively; the methodology is then evaluated in Sec.~\ref{sec:exp}.
 %In Sec.~\ref{sec:rpm}, we explain how metrics based on relative performance can enable intelligent sampling. 
 Finally, in Sec.~\ref{sec:con}, we draw conclusions and discuss the applications.

\section{Related Works}
\label{sec:rel}

The problem of mapping one target linear algebra expression to a sequence of library calls is known as the Linear Algebra Mapping Problem (LAMP)~\cite{psarras2019linear} \ac{and it is mainly applied in auto-tuning High-Performance Computing applications~\cite{Balaprakash2018AutotuningIH}}; typical problem instances have many mathematically equivalent solutions, and high-level languages and environments such as Matlab~\cite{MatlabOTB}, Julia \cite{julia} etc., ideally should select the best one. However, it has been shown that most of these languages choose algorithms that are sub-optimal in terms of performance~\cite{psarras2019linear,barthels2019linnea}. A general approach to identify the best candidate is by ranking the solution algorithms according to their predicted performance. For linear algebra computations, the most common performance metric to be used as performance predictor is the FLOP count, even though it is known that the number of FLOPs is not always a direct indicator of the fastest code~\cite{barthels2019linnea}, especially when the computation is bandwidth-bound or executed in parallel. For selected bandwidth-bound operations, Iakymchuk et al. developed analytical performance models based on memory access patterns~\cite{iakymchuk2012modeling,iakymchuk2011execution}; while those models capture the program execution accurately, their construction requests not only a deep understanding of the processor, but also of the details of the implementation. 
%\as{I cited elmars micro benchmark paper below}

Performance metrics that are a summary of execution times (such as minimum, median etc.) lack in consistency when the measurements of the programs are repeated;  this is due to system noise~\cite{hoefler2010characterizing}, cache effects~\cite{peise2014cache}, behavior of collective communications \cite{agarwal2005impact} etc., and it is not realistic to eliminate the performance variations entirely~\cite{trackingPerfVariation}. The distribution of execution times obtained by repeated measurements of a program is known to lack in textbook statistical behaviors and hence specialized methods to quantify performance have been developed~\cite{robustbenchmarking,statiscalperfCompare,hoefler2010loggopsim,davidWaitStates}. However, approximating statistical distributions require executing the algorithms several times. 

The performance of an algorithm can be predicted using regression or machine learning based methods; this requires careful formulation of an underlying problem. A wide body of significant work has been done in this direction for more than a decade~\cite{peise2014performance,regressionScalability,barve2019fecbench,Jessup2016PerformanceBasedNS}. 
\ac{Peise et al in~\cite{peise2014performance}} creates a prediction model for individual BLAS calls and estimates the execution time for an algorithm by composing the predictions from several BLAS calls. In~\cite{barve2019fecbench} Barve et al predicts performance to optimize resource allocation in multi-tenant systems. Barnes et al in~\cite{regressionScalability} predicts scalability of parallel algorithms. In all these approaches, measurements are designed to estimate the parameters of certain performance ``model''; once these parameters are estimated, the model can automatically predict the performance for a specific use-case and a well-defined compute environment. Porting the prediction model for a different compute setting requires running the measurements and re-parametrizing the model again.

Online performance prediction can be used to tackle the sensitivities to compute environments; the dynamic nature of performance is encoded through real-time update of the model parameters via algorithm comparisons~\cite{reiji,adaptiveOnline,comparingKrylov}. In~\cite{reiji}, Reiji Suda selects the best algorithm by solving a ``multi-armed bandit problem''. Here, an application program calls a library function iteratively and there are several equivalent algorithms for the library routine. Candidate algorithms are compared among each other during the course of the iteration and a Bayesian model is updated to minimize the total execution time.

Performance metrics based on algorithm comparisons (or discrimination) are central to developing a class of performance models based on reinforcement learning. The multi-armed bandit problem is a reinforcement learning problem. In such approaches, the model parameters are updated using a reward function, which is formulated based on algorithm discriminations; faster algorithms gets higher rewards and the slower algorithms are penalized. Therefore, in this paper, we develop and analyse a metric based on relative performance, which is estimated by clustering the algorithms into performance classes. This metric can be used to define rewards and facilitate development of a class of practical performance models for real-time applications.  

In~\cite{paiseSankaran}, we discuss the application of relative performance in an edge computing environment, where scientific computations are split among various devices. Our relative performance metric, based on clustering methodology, can be used to select algorithms with respect to more than one criteria; for instance, from a subset of equivalently fast algorithm, one can then select an algorithm that consumes the least energy.

\section{Towards Relative Performance}
\label{sec:torel}
%The ranking methodology described in section 2 establishes a notion of distance (in terms of performance) between the algorithms in $\mathcal{A}$.
Fundamentally, for every algorithm in $\mathcal{A}$ we want to compute a score that indicates its chance of being the
fastest. As an example, consider the distributions of execution times shown in Fig.~\ref{fig:1} for three different algorithms $\mathcal{A} := \{ $``Red", ``Blue", ``Yellow"$\}$ that solve the Ordinary Least Square problem:\footnote{The pseudocode of solution algorithms are shown in Appendix A. For the sake of
  explanation, we only consider three solution algorithms; for this problem, Linnea generates many more than three~\cite{barthels2019linnea}.} $(X^TX)^{-1}X^{T}y$ where $X \in \mathbb{R}^{m \times n}$ and $y \in \mathbb{R}^{n}$. 
The distributions in the example were obtained by measuring the execution time of every algorithm $\mathbf{alg}_j \in
\mathcal{A}$,  500 times\footnote{In Sec. \ref{sec:exp} B, we show that our methodology produces reliable results also
  for smaller values of $N$.}; let $\mathbf{t_j} \in \mathbb{R}^{500}$ be the distribution of execution times for algorithm $\mathbf{alg}_j$.
In order to ensure that the time measurements are unbiased of system noise, we adopted the following procedure: Let the set of all executions of the three algorithms be $\mathcal{E} = \mathbf{e}_1 \oplus \mathbf{e}_2 \oplus \mathbf{e}_3$, where $\mathbf{e}_j \in \{(\mathbf{alg}_j)_1 \dots (\mathbf{alg}_j)_{500}\}$ is the set of 500 executions of $\mathbf{alg}_j$ and  $\oplus$
is a concatenation operation. The set $\mathcal{E}$ was shuffled before the executions were timed and the measurements $\mathbf{t}_j$ were obtained. 
%Every execution $e \in \mathcal{E}$ was run twice and only the second measurement was considered after the cache was trashed.

A ``straightforward'' approach to relatively score algorithms consists in finding the minimum (or median, or mean) execution time of every algorithm $\mathbf{alg}_j \in \mathcal{A}$ from $N=500$ measurements and then use that to construct a ranking~\cite{peise2012performance}. By doing this, every algorithm
  would obtain a unique rank. If we choose to assign the algorithm with rank 1 to the set of fastest algorithm
  $\mathcal{F}$, then only one algorithm (Yellow or Blue) could be assigned to $\mathcal{F}$, although the distributions of Yellow and
  Blue are nearly identical. This will lead to inconsistent rankings when all the executions $\mathcal{E}$ are
  repeated, unless there is a methodology to assign rank 1 to both Yellow and Blue algorithms. Reproducibility of rankings is essential in order to derive mathematically sound performance
  predictions. In order to ensure reproducibility of relative performance estimates, both Yellow and Blue should be assigned to $\mathcal{F}$. 
  To this end, one could choose the $k$-best algorithms\cite{kbest-kadioglu2011algorithm} with $k=2$, and  assign both
  Yellow and Blue to $\mathcal{F}$. However, the fixed value $k=2$  might not be
  suitable for other scenarios in which either only one algorithm or more than two algorithms are superior to the rest.

\paragraph*{\textbf{Bootstrapping}} Bootstrapping\cite{bootstrap} is a common procedure to summarize statistics from an arbitrary distribution. Here, the ``straightforward'' approach is repeated $Rep$ times, and for each iteration, $K < N$ measurements are sampled from the original $N$ measurements and rankings are constructed. If an algorithm $\mathbf{alg_j}$ obtains rank 1 in at least one out of $Rep$ iterations, it is assigned to $\mathcal{F}$, and will
 receive the score of  $c/Rep$, where $c \le Rep$ is the number of times $\mathbf{alg_j}$ obtains rank 1. The steps for
 this approach are shown in Procedure \ref{alg:fa}. For a given set of algorithms $
 \mathbf{alg}_1,\mathbf{alg}_2 ,\dots, \mathbf{alg}_p\in \mathcal{A}$ for which the execution times $
 \mathbf{t}_1,\mathbf{t}_2 ,\dots, \mathbf{t}_p\in \mathbb{R}^{N}$ are measured $N$ times each,
 Procedure \ref{alg:fa} returns the subset $\mathcal{F}  \subseteq \mathcal{A}$ of fastest algorithms $\mathbf{f}_1, \mathbf{f}_2, \dots, \mathbf{f}_q \in  \mathcal{F}$ ($q \le p$), each with an associated score $c_1,c_2,\dots,c_q$.
 As an example, when this procedure is applied to the distributions in Fig.~\ref{fig:1} (with $Rep=100$ and a sample size $K=5$), both Yellow and Blue are assigned to $\mathcal{F}$, with scores 0.55 and 0.45 respectively. 

\begin{algorithm}
	\caption{ Get$\mathcal{F}$$(\mathcal{A}, Rep, K)$ }
	\label{alg:fa}
	\hspace*{\algorithmicindent} \textbf{Input: } $ \mathbf{alg}_1,\mathbf{alg}_2 ,\dots, \mathbf{alg}_p\in \mathcal{A} \qquad K,Reps \in \mathbb{Z}^{+} $  \\
	\hspace*{\algorithmicindent} \textbf{Output: } $ (\mathbf{f}_1,c_1), (\mathbf{f}_2, c_2), \dots, (\mathbf{f}_q,c_q) \in \mathcal{F} \times \mathbb{R}  $
	\begin{algorithmic}[1] 
		\State $p = |\mathcal{A}|$ 
		\For{i = 1, $\dots$, $p$}
		\State $\mathbf{t}_i = get\_measurements(\mathbf{alg}_i)$ \Comment{$\mathbf{t_i} \in \mathbb{R}^{N}$}
		\EndFor
		\State $\mathbf{a}\leftarrow [ \quad ]$ \Comment{Initialize empty lists}
		\State $\mathbf{f} \leftarrow [ \quad ]$ 
		\For{i = 1, $\dots$, $Rep$}
		\For{j = 1, $\dots$, $p$}
		\State $\mathbf{\tilde{t}_j} \leftarrow $  sample $K $ values from $\mathbf{t_j}$ \Comment{$\mathbf{\tilde{t}_j} \in \mathbb{R}^{K}$}
		\EndFor
		\State $ \mathbf{a}_i \leftarrow algorithm( \underset{j}{\mathrm{argmin}}(\min(\mathbf{\tilde{t}_j})))$ \Comment{$\mathbf{a}_i \in \mathcal{F}$}
		\EndFor
		\State $\mathbf{f} \leftarrow $ select unique algorithms in $\mathbf{a}$ \Comment{$|\mathbf{f}| \le |\mathbf{a}|$}  
		\State $q = |\mathbf{f}|$
		\For{i = 1, $\dots$,$ q$ }
        \State $c_i \leftarrow$ number of occurrences of $\mathbf{f}_i$ in $\mathbf{a}$
		\State $c_i \leftarrow c_i/Rep$
		\EndFor
		\State return $ (\mathbf{f}_1,c_1), (\mathbf{f}_2, c_2), \dots, (\mathbf{f}_q,c_q) $ 
              \end{algorithmic}
%\p{flush comments to the right margin (maybe with hfill?)}
\end{algorithm}

%Let $\mathcal{A}(x)$ be the set of all solution algorithms for certain linear algebra expression $x$. 
However, in practical applications, the execution times of algorithms are typically
measured only a few times (likely much fewer than 500 times, possibly just once), and therefore, only a snapshot of the distributions
shown in Fig.~\ref{fig:1} will be available. Procedure \ref{alg:fa} does not account for the uncertainty in measurement data in capturing the true distributions. 
%The scores obtained by Procedure \ref{alg:fa} still does have a meaningful translation to the indicate the distance between the algorithms; ideally, the scored obtained for Red and Blue should be comparable. This is because only one algorithm is assigned to $\mathcal{F}$ in every iteration. 
The recommended approach is to use tests for significant differences (such as Kruskal-Walis
ANOVA~\cite{hoefler2015scientific}) between pairs of algorithms, and to merge the ranks of algorithms if there is not enough
evidence of one algorithm dominating over the other. In this paper, we incorporate the test for significant differences in
Procedure \ref{alg:fa}  by introducing a
three-way comparison function. 

\section{Methodology}
\label{sec:met}

For each algorithm $\mathbf{alg}_1, \dots, \mathbf{alg}_p \in \mathcal{A}$, its execution time is measured $N$ times using the measurement strategy described in Section \ref{sec:torel}. Let $\mathbf{t}_j \in \mathbb{R}^{N}$ be the array of time measurements for algorithm $\mathbf{alg_j}$. We do not make any assumptions about the nature of the distributions of timings. In order to compute the relative score for an algorithm, we first cluster all the algorithms in $\mathcal{A}$ into performance classes. Let $r_1, r_2, \dots,r_w \in \mathbb{Z}^{+}$ be the ranks of $w$ performance classes. The number of performance classes\footnote{The number of performance classes $w$ is determined dynamically and need not be specified by the user.} (or ranks)  $w $ is lesser than or equal to the total number of algorithms $p$ ($w \le p$), as more than one
algorithm can be assigned with the same rank. For illustration, let us consider an example of clustering four algorithms
$\mathbf{alg}_1,\mathbf{alg}_2, \mathbf{alg}_3, \mathbf{alg}_4$  ($p=4$) into performance classes.  Significant overlap in the distributions are observed between algorithms $\mathbf{alg}_2$, $\mathbf{alg}_4$ and  $\mathbf{alg}_1$, $\mathbf{alg}_3$; thus, the algorithms should be ranked into two performance classes ($w=2$). We represent the outcome of this procedure as a sequence set with tuples $(\mathbf{alg}_j,r_j)$, where $r_j$ is the rank assigned to $\mathbf{alg}_j$ and the sequence set for this illustration should result as $\langle (\mathbf{alg}_2,1),
(\mathbf{alg}_4,1), (\mathbf{alg}_1,2), (\mathbf{alg}_3,2) \rangle$; where an algorithm with a lesser rank is better. 
  All the algorithms with rank 1 are assigned to the set of fastest algorithms $\mathcal{F}$. To this end, we first sort the algorithms in $\mathcal{A}$ using a three-way comparison function.

%  As the number of time measurements $N$ decreases, the assignment of ranks need not be deterministic; that is, the ranks can change when the ``first step'' is repeated, even for the same time measurements. Hence, as a ``second step", we bootstrap and repeat the ``first step" of the procedure $Rep$ times and all the algorithms that obtained rank 1 at least once are assigned to $\mathcal{F}$ and the relative scores are computed. Through the repetition of the ``first step'', we extract as much information as possible from the distributions, which leads to stable relative score estimates in the ``second step''; that is, the deviations in the scores would be minimal, even when the time measurements are repeated. 

The comparison of the distributions of execution times of any two algorithms $\mathbf{t_i},\mathbf{t_j} \in \mathbb{R}^{N}$ is indicated in Procedure \ref{alg:compare}.
%\p{``method'' is not a common word for this. ``Algorithm'', ``Procedure'', ``Test'', ... Later you call it ``function'', which is also ok}
%
The comparison yields one of three outcomes:
\begin{itemize}
\item Outcome A: algorithm $\mathbf{alg_i}$ is faster than $\mathbf{alg_j}$ $(<)$
\item Outcome B: algorithm  $\mathbf{alg_i}$ is as good as $\mathbf{alg_j}$ $(\sim)$
\item Outcome C: algorithm  $\mathbf{alg_i}$ is slower than $\mathbf{alg_j}$ $(>)$
\end{itemize}
Outcomes A and C imply that there are noticeable differences between the distributions; outcome B implies that the
distributions are performance-wise equivalent. 
\begin{algorithm}
	\caption{CompareAlgs $(\mathbf{alg}_i, \mathbf{alg}_j, threshold,M,K)$ }
	\label{alg:compare}
	\hspace*{\algorithmicindent} \textbf{Inp: } $ \mathbf{alg}_i, \mathbf{alg}_j \in \mathcal{A} \qquad threshold \in [0.5,1]$ \\ 
	\hspace*{\algorithmicindent} \hspace*{\algorithmicindent}  $  \quad M,K \in \mathbb{Z}^{+} $\\ 
	\hspace*{\algorithmicindent} \textbf{Out:} $  \{``\mathbf{alg}_i < \mathbf{alg}_j", ``\mathbf{alg}_i>\mathbf{alg}_j", ``\mathbf{alg}_i\sim \mathbf{alg}_j" \}$
	\begin{algorithmic}[1] 
		\State $\mathbf{t_i} = get\_measurements(\mathbf{alg}_i)$ \Comment{$\mathbf{t_i} \in \mathbb{R}^{N}$}
		\State $\mathbf{t_j} = get\_measurements(\mathbf{alg}_j)$ \Comment{$\mathbf{t_j} \in \mathbb{R}^{N}$}
		\State $p \leftarrow 0$
		\For{m = 1, $\dots$, $M$}
        \State $\mathbf{\tilde{t}_i} \leftarrow$ sample $K$ values from $\mathbf{t_i}$
        \State $e_i$ = min($\mathbf{\tilde{t}_i}$)
        \State $\mathbf{\tilde{t}_j} \leftarrow$ sample $K$ values from $\mathbf{t_j}$
		\State $e_j$ = min($\mathbf{\tilde{t}_j}$)
		\If{$e_i \le e_j$}
		\State $c=c+1$
		\EndIf
		\EndFor
		\If{$\frac{c}{M} \geq threshold$ }
		\State return ``$\mathbf{alg}_i < \mathbf{alg}_j$"
		\ElsIf{$\frac{c}{M} < 1 - threshold$}
		\State return ``$\mathbf{alg}_i>\mathbf{alg}_j$" 
		\Else
		\State return ``$\mathbf{alg}_i\sim \mathbf{alg}_j $"
		\EndIf
	\end{algorithmic}
      \end{algorithm}
%\p{The level of detail is good, but the connection is missing. ``To this end'' is not linking to anything.
%  Also, use references (``one could also use other metrics'' -- as done in [ref], [ref].}  
In Procedure \ref{alg:compare}, the distributions $\mathbf{t_i}$ and $\mathbf{t_j}$ are sampled ($\mathbf{\tilde{t}_i}
\subset \mathbf{t_i}$ and $\mathbf{\tilde{t}_j} \subset \mathbf{t_j}$) with sample size $K < N$, and the minimum
execution time from the respective samples are compared. This comparison procedure is repeated $M \ge 1$ times and the counter $c$ is updated based on the result of comparison in each iteration.
In theory, one could
also use other metrics such as mean or median execution time (instead of the minimum statistic in line 6 and 8 of Procedure \ref{alg:compare}), or even a combination of multiple statistics across different
comparisons. Peise et al. in ~\cite{peiseThesis} show that minimum and median statistics are less sensitive to
fluctuations in execution time for linear algebra computations.

Let $e_i$ and $e_j$ be the minimum execution time of samples  $\mathbf{\tilde{t}_i}$ and $\mathbf{\tilde{t}_j}$,
respectively. The probability that $e_i$ is less than $e_j$ is approximated from the results of $M$ comparisons. That
is, if $e_i$ is less than $e_j$ in $c$ out of $M$ comparisons, then the empirical probability $P[e_i \le e_j]$ is
$c/M$. The outcome of the CompareAlgs function (Procedure \ref{alg:compare}) is determined by comparing the empirical probability against certain $threshold$ (see lines 11 - 16 in Procedure \ref{alg:compare}).  The outcome of CompareAlgs function is not entirely
deterministic: If $c/M \to threshold$, then  $\mathbf{alg_i}$ approaches the ``state of being faster'' than $\mathbf{alg_j}$ and
the outcome can tip over A ($<$) or B ($\sim$) when the CompareAlgs function is repeated (even for the same input distributions). As a consequence, the comparisons need not be transitive;  $\mathbf{alg_1} < \mathbf{alg_2}$ and
$\mathbf{alg_2} < \mathbf{alg_3} $ does not imply $\mathbf{alg_1} < \mathbf{alg_3}$.

\textbf{\textit{Effect of }$threshold$:} Valid values of $threshold$ for Procedure \ref{alg:compare} lie in the range $[0.5,1]$. When $threshold \to 0.5$, the outcome B ($\sim$) becomes
less and less likely (and impossible for $threshold=0.5$), while for $threshold \to 1$, the conditions for outcomes A ($<$)
and C ($>$) becomes stricter and outcome B become more and more likely. 

\textbf{\textit{Effect of} $K$:} As $K \to N$, the minimum of  the samples ($e_i, e_j$) approximates the
minimum execution time ($m_i, m_j$) of the distributions $\mathbf{t_i}$ and $\mathbf{t_j}$ respectively, and the
result of comparison $e_i \le e_j $ (line 9 of Procedure \ref{alg:compare}) becomes more and more deterministic. As a consequence, $c/M$ tends to either 0 or 1 and outcome B
($\mathbf{alg}_i \sim \mathbf{alg}_j$) becomes less and less likely. When $K=N$,  outcome B becomes impossible and this invalidates the advantages of bootstrapping. When $K \to 1$, the minimum estimates of samples ($e_i, e_j$) point to an instance of execution time $t_i, t_j \in \mathbb{R}$ from the distribution. Then, outcome B becomes more likely even for marginal overlap of distributions (as in Fig \ref{fig:diff}).
% especially when $threshold$ is high or $M$ is small.%

\textbf{\textit{Effect of} $M$:} When $M=1$, the $threshold$ does not affect the outcome of comparison and the outcome B becomes impossible. As $M \to N$, the accuracy of the empirical probability $P[e_{i} \le e_{j}]$ for a given $K$ increases.

\begin{figure}
	\includegraphics[width=0.5\textwidth]{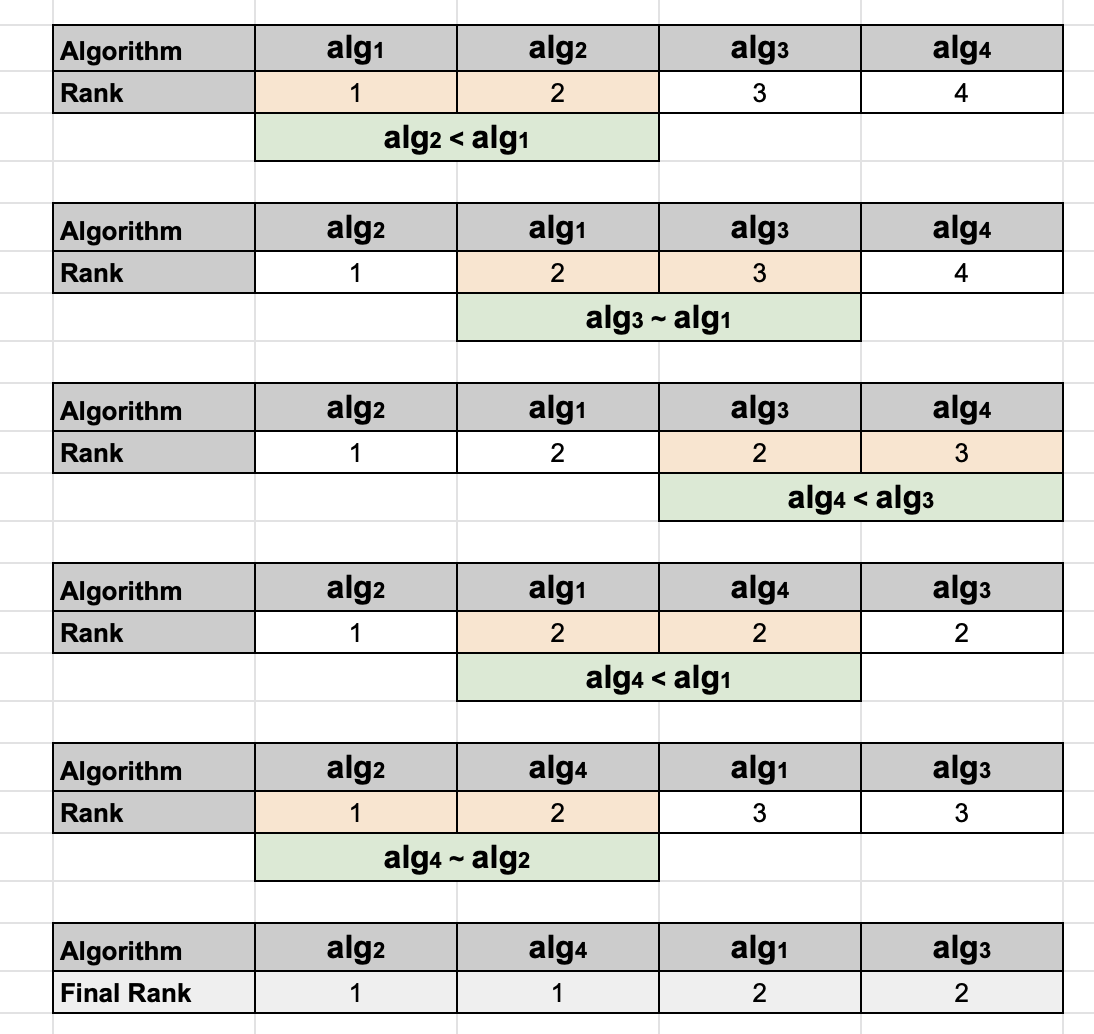}
	\caption{Bubble Sort with Compare function}
	\label{fig:sort}
\end{figure}

\paragraph*{\textbf{Sorting procedure}} The CompareAlgs function (Procedure \ref{alg:compare}) is used in Procedure \ref{alg:sort} to sort the algorithms $\mathbf{alg}_1,\mathbf{alg}_2, \dots, \mathbf{alg}_p \in \mathcal{A}$ and assign them to one of the performance classes $r_1, r_2, \dots, r_w$. Initially, the number of performance classes $w$ is same as the number of algorithms $p$.
The outcome of Procedure \ref{alg:sort} is represented as a sequence set $\langle (\mathbf{alg}_{s[1]},r_1), \dots, (\mathbf{alg}_{s[p]},r_w) \rangle$; for instance, if $\langle (\mathbf{alg}_{2},1), (\mathbf{alg}_{4},1) , (\mathbf{alg}_{1},2), (\mathbf{alg}_{3},2) \rangle$ is the returned sequence, $\mathbf{s}_1$ (or $\mathbf{s}[1]$) is the index of the first (and best) algorithm, which is 2, and this algorithm obtains rank 1. $\mathbf{s}_2$ is the second index, which is 4 with rank 1 and so on. To this end, a swapping procedure that adapts bubble-sort\cite{bubblesort} with three-way comparisons is used (the bubble-sort algorithm is used only to demonstrate the idea behind the methodology; therefore it is not optimized for performance). Procedure \ref{alg:sort} first creates an initial sequence set with all the algorithms in $\mathcal{A}$ assigned with distinct consecutive ranks $\langle (\mathbf{alg}_1,1), \dots, (\mathbf{alg}_p,p) \rangle$.
%Procedure~\ref{alg:sort} illustrates the swapping procedure that adapts bubble-sort\cite{bubblesort} with three-way comparisons (the bubble-sort algorithm is used only to demonstrate the idea behind the methodology; therefore it is not optimized for performance).
\begin{algorithm}
	\caption{SortAlgs $(\mathcal{A}, thresh,M,K)$ }
	\label{alg:sort}
	\hspace*{\algorithmicindent} \textbf{Input: } $ \mathbf{alg}_1,\mathbf{alg}_2,\dots,\mathbf{alg}_p \in \mathcal{A}$ \\
	\hspace*{\algorithmicindent} \hspace*{\algorithmicindent} $ \quad thresh \in [0.5,1] \quad M,K \in \mathbb{Z}^{+}$ \\
	\hspace*{\algorithmicindent} \textbf{Output: } $ \langle (\mathbf{alg}_{s[1]},r_1), (\mathbf{alg}_{s[2]}, r_2), \dots, (\mathbf{alg}_{s[p]},r_w) \rangle $
	\begin{algorithmic}[1] 
		\State Initialize w $\leftarrow$ p
		%\State Initialize $r_1,\dots,r_w$ with consecutive ranks 
		\For{i = 1, $\dots$, p}
		\State  Initialize $\mathbf{r}_i \leftarrow i$ \Comment{Initialize Alg rank}
		\State  Initialize $\mathbf{s}_i \leftarrow i$ \Comment{Initialize Alg Index}
		\EndFor
		%\State Initialize $\mathbf{a}_{s[j]} \leftarrow \mathbf{a}_j \quad \forall  j \in \{1,2,..,p\}$
		\For{i = 1, $\dots$, p}
		\For{j = 0, $\dots$, p-i-1}
		\State ret = CompareAlgs($\mathbf{alg}_{s[j]}, \mathbf{alg}_{s[j+1]}, thresh, M, K$)
		\If{$\mathbf{alg}_{s[j+1]}$ is faster than $\mathbf{alg}_{s[j]}$}
		\State Swap values of $\mathbf{s}_{j}$ and $\mathbf{s}_{j+1}$ \label{lst:swap}
		\If{$r_{j+1} = r_j$} \label{lst:h1}
		\If{$r_{j-1} \ne r_j$ or $j=0$}
		\State Increment ranks $r_{j+1}, \dots, r_p$ by 1 \label{lst:h2}
		\EndIf
		\Else
		\If{$r_{j-1} = r_j$ and $j\ne0$} \label{lst:gg1}
		\State Decrement ranks $r_{j+1}, \dots, r_p$ by 1 \label{lst:gg2}
		\EndIf
		\EndIf
		\ElsIf{$\mathbf{alg}_{s[j+1]}$ is as good as $\mathbf{alg}_{s[j]}$} \label{lst:ag1}
		\If{$r_{j+1} \ne r_j$}
		\State Decrement ranks $r_{j+1}, \dots, r_p$ by 1 \label{lst:ag2}
		\EndIf
		\ElsIf{$\mathbf{alg}_{s[j]}$ is faster than $\mathbf{alg}_{s[j+1]}$}
		\State Leave the ranks as they are
		\EndIf		
		\EndFor
		\EndFor
		\State return $\langle (\mathbf{alg}_{s[1]},r_1), \dots, (\mathbf{alg}_{s[p]},r_w) \rangle$
              \end{algorithmic}
            \end{algorithm}
 Starting from the first element of the initial sequence, bubble sort compares adjacent algorithms and swaps
 indices if an algorithm occurring later in the sequence $\mathbf{alg}_{s[j+1]}$ is faster (according to Procedure \ref{alg:compare}) than the previous algorithm
 $\mathbf{alg}_{s[j]}$, and then ranks are updated. When the comparison of two algorithms results to be as good as each other, both are assigned with the same rank, but their indices are not swapped. In order to illustrate the rank update rules in detail, consider the example in Fig
 \ref{fig:sort}, which shows the intermediate steps while sorting the distributions of four algorithms $\mathbf{alg}_1,
 \mathbf{alg}_2, \mathbf{alg}_3,\mathbf{alg}_4$ (initialized with ranks 1,2,3,4 respectively). All possible update rules that
 one might encounter appear in one of the intermediate steps of this example.
%If an algorithm $\tilde{s}_{j+1}$ is better than its previous algorithm in sequence $\tilde{s}_j$, then the position index of the two algorithms are swapped.

\begin{enumerate}
	\item 
	      \textit{\textbf{Both indices and ranks are swapped} :}
          In the first pass of bubble sort, pair-wise comparison of adjacent algorithms are done starting from the first
          element in the sequence. Currently, the sequence is $\langle \mathbf{alg}_1, \mathbf{alg}_2, \mathbf{alg}_3, \mathbf{alg}_4 \rangle$.
          As a first step, algorithms $\mathbf{alg}_1$ and $\mathbf{alg}_2$ are compared, and 
          $\mathbf{alg}_2$ ends up being faster. As the slower algorithm should be shifted towards the end of the sequence, $\mathbf{alg}_1$ and $\mathbf{alg}_2$ swap positions  (line \ref{lst:swap} in
          Procedure \ref{alg:sort}). Since all the algorithms still have unique ranks, $\mathbf{alg}_2$ and
          $\mathbf{alg}_1$ also exchange their ranks, and no special rules for updating ranks are applied.
          So, $\mathbf{alg}_2$ and $\mathbf{alg}_1$ receive rank 1 and 2, respectively.
	
	\item \textit{\textbf{Indices are not swapped and the ranks are merged}:} Next, algorithm $\mathbf{alg}_1$ is compared with its successor $\mathbf{alg}_3$;
          since they are just as good as one another, 
          no swap takes place. Now, the rank of $\mathbf{alg}_3$ should also indicate that it is as good as $\mathbf{alg}_1$;
          so $\mathbf{alg}_3$ is given the same rank as $\mathbf{alg}_1$ 
            and the rank of $\mathbf{alg}_4$ is corrected by
         decrementing by 1. (line \ref{lst:ag1}-\ref{lst:ag2} in Procedure \ref{alg:sort}). Hence $\mathbf{alg}_1$ and $\mathbf{alg}_3$ have rank 2, and $\mathbf{alg}_4$ is corrected to rank 3.
	
       \item
         \textit{\textbf{Indices are swapped and the ranks are merged}:} 
         In the last comparison of the first sweep of bubble sort, Algorithm $\mathbf{alg}_4$ results to be faster than $\mathbf{alg}_3$, so their positions are
          swapped. However, ranks cannot be exchanged the same way it happened in step 1,
          as $\mathbf{alg}_3$ would receive rank 3, even though it was evaluated to be
          as good as $\mathbf{alg}_1$, which has rank 2 (recall from step 2: $\mathbf{alg}_3 \sim \mathbf{alg}_1$). So
          instead of changing the rank of $\mathbf{alg}_3$, the rank of $\mathbf{alg}_4$ is improved (line \ref{lst:gg1} -
          \ref{lst:gg2} in Procedure \ref{alg:sort}). Thus  $\mathbf{alg}_1$,  $\mathbf{alg}_4$, and  $\mathbf{alg}_3$ are
          given rank 2, but $\mathbf{alg}_4$ is now earlier in the sequence than $\mathbf{alg}_3$. This indicates that there
          is an inherent ordering even when algorithms share the same rank. Even though $\mathbf{alg}_4$ gets the same rank as $\mathbf{alg}_3$ despite resulting to be faster than the latter, it still has chances to be assigned with better rank as the sort procedure progress.
          Every pass of bubble sort pushes the slowest algorithm to the end of the sequence.
          At this point, the sequence is $\langle \mathbf{alg}_2, \mathbf{alg}_1, \mathbf{alg}_4, \mathbf{alg}_3 \rangle$.
		
	\item
          \textit{\textbf{Swapping indices with algorithms having same rank}:} In the second pass of bubble sort, the pair-wise comparison of adjacent algorithms, except the ``right-most" algorithm in sequence, is evaluated (note that the right-most algorithm can still have its rank updated depending upon the results of comparisons of algorithms occurring earlier in the sequence). The first two algorithms $\mathbf{alg}_2$ and
          $\mathbf{alg}_1$ were already compared in Step 1. So now, the next comparison is $\mathbf{alg}_1$ vs.~$\mathbf{alg}_4$.
          Algorithm $\mathbf{alg}_4$ results to be faster than $\mathbf{alg}_1$, so the algorithms are swapped as usual. However, the
          algorithms currently share the same rank. Since $\mathbf{alg}_4$ reached the top of its performance
          class---having defeated all the other algorithms with the same rank---it should now get a higher rank than the
          other algorithms in its class. Therefore, the rank of $\mathbf{alg}_1$ and $\mathbf{alg}_3$ is incremented by 1
          (line \ref{lst:h1}-\ref{lst:h2} in Procedure \ref{alg:sort}): $\mathbf{alg}_4$ stays at rank 2, and $\mathbf{alg}_1$ and $\mathbf{alg}_3$ receive rank 3. This completes the second pass of bubble sort and the two slowest algorithms have been pushed to the right.
	
	\item
          \textit{\textbf{Indices are not swapped and the ranks are merged:}} (\textit{This is the same rule applied in Step 2}). In the third and final pass,
          we again start from the first element on the left of the sequence and continue the pair-wise comparisons until the third last
          element. This leaves only one comparison to be done, $\mathbf{alg}_4$ vs.~$\mathbf{alg}_2$.
          Algorithm $\mathbf{alg}_4$ is evaluated to be as good as $\mathbf{alg}_2$, so both are given the same rank and the positions are not swapped. The ranks of algorithms occurring later than $\mathbf{alg}_4$ in the sequence are decremented by 1. Thus,
          the final sequence is $\langle \mathbf{alg}_2, \mathbf{alg}_4, \mathbf{alg}_1, \mathbf{alg}_3 \rangle$. Algorithms $\mathbf{alg}_2$ and $\mathbf{alg}_4$ obtain rank 1, and $\mathbf{alg}_1$ and $\mathbf{alg}_3$ obtain rank 2.
\end{enumerate}

All the algorithms with rank 1 are assigned to the set of fastest algorithms $\mathcal{F}$. From the above illustration,
$\mathcal{F} \leftarrow \{\mathbf{alg}_2, \mathbf{alg}_4\}$. The results of the Compare Function (Procedure \ref{alg:compare}) are not entirely deterministic due to the non-transitivity of comparisons. As a consequence, the final ranks obtained by Procedure \ref{alg:sort} are also not entirely deterministic.
For instance, in step 5, algorithm $\mathbf{alg}_4$ could be right at the threshold  of being better than $\mathbf{alg}_2$;
that is, if $\mathbf{alg}_4$ were estimated to be better than $\mathbf{alg}_2$ once in every two runs of Procedure \ref{alg:sort}, then fifty percent of the times
$\mathbf{alg}_2$ would be pushed to rank 2 and would not be assigned to $\mathcal{F}$. To address this, the SortAlgs function
is repeated
$Rep$ times and in each iteration, all the algorithms that received rank 1 are accumulated (in the list `$\mathbf{a}$' in Procedure \ref{alg:f}). Then, we follow the steps similar to lines 10-14 in Procedure \ref{alg:fa} to compute the relative score. If an
algorithm $\mathbf{alg}_j$ was assigned rank 1 in $c$ out of $Rep$ iterations, then $\mathbf{alg}_j$ would appear $c$ times in
the list `$\mathbf{a}$'. Then, the relative confidence (or relative score) for $\mathbf{alg}_j$ is $c/Rep$. 
%Notice that no bootstrapping is done in each of the $T$ iterations (as it was done in Procedure \ref{alg:fa}). The Sort function is repeated only to address the randomness in the final rankings and not to extract more information from the distribution (which is the objective of Bootstrapping that was addressed in the Compare function). 
The modified version of Procedure \ref{alg:fa} is shown in Procedure \ref{alg:f}, which returns a set of fastest algorithms $\mathcal{F}$ (all the unique occurrences in $\mathbf{a}$)
and their corresponding relative scores\footnote{Code: https://github.com/as641651/Relative-Performance}. Since we are only interested in the fastest algorithms, all the other
candidates that were not assigned rank 1 even once are given a relative score of 0. For the illustration in Figure 1, if $\mathbf{alg}_4 < \mathbf{alg}_2$ in approximately one out of two executions of SortAlgs function, $\mathbf{alg}_2$ would get a relative score of 0.5 and $\mathbf{alg}_4$ would get 1.0.  $\mathbf{alg}_1$ and $\mathbf{alg}_3$ would be given relative scores 0. \textbf{A higher relative score implies a greater confidence of being one of the fastest algorithms; that is, the result can be reproduced with a higher likelihood when all the time measurements are repeated, irrespective of fluctuations in system noise, but for a given operation setting} (computing architecture, operating system and run time settings).

\begin{algorithm}
	\caption{ Get$\mathcal{F}$$(\mathcal{A}, Rep, thresh, M, K)$ }
	\label{alg:f}
	\hspace*{\algorithmicindent} \textbf{Input: } $ \mathbf{alg_1},\mathbf{alg_2} ,\dots, \mathbf{alg_p}\in \mathcal{A}$ \\
	 \hspace*{\algorithmicindent} \hspace*{\algorithmicindent} $ \quad thresh \in [0.5,1] \quad M,K,Reps \in \mathbb{Z}^{+}$ \\
	\hspace*{\algorithmicindent} \textbf{Output: } $ (\mathbf{f}_1,c_1), (\mathbf{f}_2, c_2), \dots, (\mathbf{f}_q,c_q) \in \mathcal{F} \times \mathbb{R}  $
	\begin{algorithmic}[1] 
		\State $\mathbf{a}\leftarrow [ \quad ]$ \Comment{Initialize empty lists}
		\State $\mathbf{f} \leftarrow [ \quad ]$ 
		\For{i = 1, $\dots$, $Rep$}
		\State SortAlgs$(\mathcal{A}, thresh, M, K)$ 
%                \p{still struggling with this Sort. It would really be better if the name was immediately identified as
%                  the sort from Procedure 3 -- I think we should have a slightly different name. SortBLAH. Then, we
%                  would not struggle to denote the output, which right now is superconvoluted. Now that I get it, I feel
%                  we need to replace $[\langle (\mathbf{a}_j, r_j) | \forall j \rangle ]$ with something easy to
%                  digest. Maybe we don't need output at all, just:\\
%                  $Sort(\mathcal{A})$\\
%                  Select algs with rank 1\\
%                 }
% \State $\tilde{\mathcal{F}} \leftarrow (\mathbf{s}_j | r_j = 1 )$ \Comment{select all algorithms with rank 1}
		\State $\tilde{\mathbf{a}} \leftarrow$ select algorithms with rank 1
%                \p{should we use an action word such as "select( ... ) such that .. " ?}
%                \p{what are $r_j$?}
		\State append $\tilde{\mathbf{a}}$ to the list $\mathbf{a} $
		\EndFor
		\State $\mathbf{f} \leftarrow $ select unique algorithms in $\mathbf{a}$ \Comment{$|\mathbf{f}| \le |\mathbf{a}|$}  
		\State $q = |\mathbf{f}|$
		\For{i = 1, $\dots$,$ q$ }
		\State $c_i \leftarrow$ number of occurrences of $\mathbf{f}_i$ in $\mathbf{a}$
		\State $c_i \leftarrow c_i/Rep$
		\EndFor
		\State return $ (\mathbf{f}_1,c_1), (\mathbf{f}_2, c_2), \dots, (\mathbf{f}_q,c_q) $ 
              \end{algorithmic}
              % \p{same comments apply. Do not use L.append, loops with "for i", syntax, ...}
\end{algorithm}

\section{Experiments}
\label{sec:exp}
\begin{figure}
	\centering
	\begin{subfigure}[b]{0.5\textwidth}
		\includegraphics[width=1\linewidth]{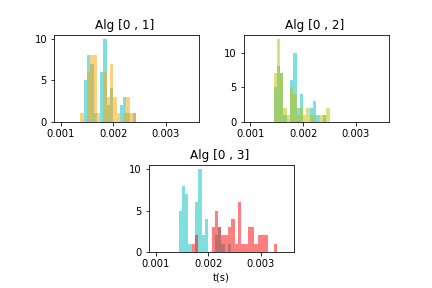}
		\caption{Setting 1 : Number of threads is set to 24}
		\label{fig:Ng1}
	\end{subfigure}
	
	\begin{subfigure}[b]{0.5\textwidth}
		\includegraphics[width=1\linewidth]{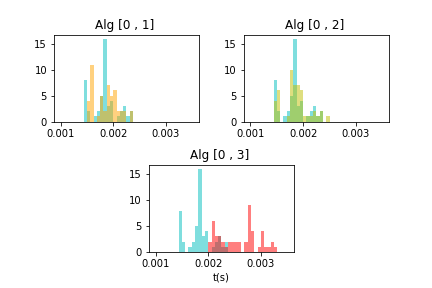}
		\caption{Setting 2 : Threads are randomly chosen between 20 - 24}
		\label{fig:Ng2} 
	\end{subfigure}
	\caption{Fifty measurements of algorithms 1,2,3 compared against algorithm 0.  Algorithm 0: Blue;  Algorithm 1: Orange; Algorithm 2: Yellow; Algorithm 3: Red.}
	\label{fig:d}
\end{figure}

The relative score can serve as a metric to discriminate an algorithm from the faster algorithms in $\mathcal{A}$. In this section, we validate the correctness of relative scores calculated using our methodology.
For our experiments, we consider four solution algorithms $\mathbf{alg}_0,
\mathbf{alg}_1, \mathbf{alg}_2, \mathbf{alg}_3 \in \mathcal{A}$ for the Ordinary Least Square problem $(X^TX)^{-1}X^{T}y$ where $X \in \mathbb{R}^{m \times n}$ and $y \in \mathbb{R}^{n}$
(pseudocode in Appendix A)
% . For the given linear algebra expression with fixed matrix sizes, Linnea generates a
% family of algorithms,
coded in the Julia language~\cite{julia}. % for sequential execution.
The distribution of every algorithm consists of 50 time measurements\footnote{We
	use the measurement strategy described in Sec. \ref{sec:torel}. The measurements were taken on a dual  socket Intel
	Xeon E5-2680 v3 with 12 cores each and clock speed of 2.2 GHz running CentOS 7.7. Julia version 1.3 was linked against
	Intel MKL implementation of BLAS and LAPACK (MKL 2019 initial release).} $\mathbf{t}_0, \mathbf{t}_1, \mathbf{t}_2, \mathbf{t}_3 \in \mathbb{R}^{50}$.   
%Fig.~\ref{fig:d} presents the distributions of 50 time measurements of four equivalent algorithms $\mathbf{a}_0,
%\mathbf{a}_1, \mathbf{a}_2, \mathbf{a}_3 \in \mathbb{R}^{50}$ for the ordinary least square problem  $(X^TX)^{-1}X^{T}y$
%(pseudocode in Appendix A).
 In order to validate the consistency of the algorithms rankings to fluctuations in system noise,  measurements were repeated under two different settings. In the first setting, the number
of threads was fixed to 24; in the second setting, the number of threads for each repetition of an algorithm was randomly chosen between 20 and 24 in order to simulate noticeable fluctuations. Fig.~\ref{fig:d} presents the distributions of time measurements for both the settings; the distribution of $\mathbf{alg}_0$ is compared against the distributions of the other three algorithms.  The
distributions of $\mathbf{alg}_0, \mathbf{alg}_1, \mathbf{alg}_2$ are largely overlapping; as these algorithms perform the same number of
floating point operations but differ in the order in which they execute matrix operations. (Note that, despite similar FLOP counts,
the order of computations can still have an effect on the execution time, due to caching effects between sequence of calls within
the algorithm\cite{peise2014cache}). For all intents and purposes, as indicated by the observations from Fig.~\ref{fig:d}, 
$\mathbf{alg}_0, \mathbf{alg}_1, \mathbf{alg}_2$ are performance equivalent and we require a statistically sound metric which indicates that choosing any of these three algorithm would not make much difference. Such a metric is important for developing machine-learning based performance prediction models. 
%In order to validate our methodology, we chose an example where similar FLOP counts
%result in nearly identical distributions.
%Algorithm $\mathbf{alg}_3$ does twice the number of FLOPs than the other three algorithms and hence noticeable difference in the distribution is observed.
Recall that the relative scores returned by Procedure \ref{alg:f} indicate the chance of an algorithm to be
assigned to the set of fast algorithms $\mathcal{F}$. In this example, one would naturally expect high scores for
$\mathbf{alg}_0, \mathbf{alg}_1$, and $\mathbf{alg}_2$. We show that the assignments made to $\mathcal{F}$ are robust and invariant to system noise or fluctuation and also when the number of time measurements is reduced (from 50).

\subsection{Robustness to Fluctuations}

If algorithms are ranked according to a single performance number, such as minimum or mean
execution time, then the resulting ranking is  generally highly sensitive to small changes to the measurement data. In other words, if
new measurements were acquired, a different ranking would arise with a high likelihood.
%\p{``reproduced'' is really a bad word, as it has a different meaning associated to it.
%  Can we say instead elaborate along these lines?\\
%  we stated that if algorithms are ranked according to a single performance number (``statistic''?), such as minimum or mean
%  execution time, then the resulting ranking is highly sensitive to small changes to the input data. In other words, if
%  new data (``measurements'', or ``timings'') were acquired, it would result in a different ranking with a high likelihood.}
We illustrate the impact of such fluctuations in the ranking of algorithms by measuring the execution times under the two different settings described earlier in this section. Table \ref{tab:2} shows the distribution statistics of the algorithms $\mathbf{alg}_0, \mathbf{alg}_1, \mathbf{alg}_2,
\mathbf{alg}_3$ for setting 1 (number of threads fixed to 24) and setting 2 (number of threads randomly chosen between 20 - 24). Under setting 1, when considering the \textbf{minimum} execution time,
$\mathbf{alg}_1$ emerges as the best algorithm. However, under setting 2, $\mathbf{alg}_2$ emerges as the best algorithm and $\mathbf{alg}_1$ is significantly behind $\mathbf{alg}_0$ and $\mathbf{alg}_2$.  If \textbf{mean} execution time is considered instead of minimum, then under setting 1, $\mathbf{a}_2$ becomes the best algorithm instead of $\mathbf{alg}_1$. Hence, relying on a single performance number does not lead to robust ranking. 
\begin{table}[h!]
	\begin{center}
		\renewcommand{\arraystretch}{1.2}
		\begin{tabular}{@{}r rrr c rrr@{}}
			\toprule
			& \multicolumn{3}{c}{Setting 1} & & \multicolumn{3}{c}{Setting 2} \\
			\cmidrule{2-4} \cmidrule{6-8}
			& min & mean & std && min & mean & std \\
			\midrule
			{$\mathbf{a}_0$ \hfill }& 1.46  & 1.76  & 0.24  && 1.47  & 1.83  & 0.24  \\
			{$\mathbf{a}_1$ } & 1.44  & 1.82   & 0.25  && 1.51  & 1.91   & 0.44  \\
			{$\mathbf{a}_2$ } & 1.46  & 1.75  & 0.28  && 1.45  & 1.85  & 0.24  \\
			{$\mathbf{a}_3$ } & 1.74  & 2.61  & 0.73  && 2.05  & 2.6  & 0.36  \\
			\bottomrule
		\end{tabular}
		\caption{Time statistics for the four algorithms (in ms).}
		\label{tab:2}
	\end{center}
\end{table}

One could argue that, despite the inconsistency, the best algorithm that is identified with a single statistic is indeed
always one of the fastest algorithms ($\mathbf{alg}_0$, $\mathbf{alg}_1$ or $\mathbf{alg}_2$). However, recall that we are interested in not just one, but all algorithms that are equivalently fast. In Fig \ref{fig:d}, we notice that the distributions of $\mathbf{alg}_0$, $\mathbf{alg}_1$ and $\mathbf{alg}_2$ are largely overlapping for time measurements under both the settings; therefore, one would expect all the three algorithms to be assigned to the set of fastest algorithms $\mathcal{F}$. We consider the assignments made to $\mathcal{F}$ to be robust only if they are invariant to measurement fluctuations. 

% often in practice the
%best algorithm has to be identified without having to execute all the equivalent algorithms; for this, realistic performance models are needed. Such
%performance models are usually trained and evaluated against certain ground truth; in our example, the distributions of
%$\mathbf{a}_0$, $\mathbf{a}_1$, $\mathbf{a}_2$ are indeed nearly identical and it is essential to know the true chance
%that each algorithm has in order to be assigned to $\mathcal{F}$. To this end, the consistency of performance estimate
%is important. Furthermore,
%in Sec.~\ref{sec:con} we describe
%practical use-cases for which it is important to identify not one, but all the fast algorithms.
 
In order to further understand the assignments made to $\mathcal{F}$, let us first describe the effect of the
hyper-parameters that were introduced in our methodology (Sec.~\ref{sec:met}). Recall that, we introduced the three-way
comparison function, which in addition to the outcomes ``faster" and ``slower", also captures the ``equivalence" ($\sim$) among the two algorithms being compared. 
% we then introduced the simple bootstrapping method in Procedure \ref{alg:fa} and argued that it does not take into account the uncertainty in measurement data and therefore the test for significant difference is necessary, in which one algorithm is evaluated to be faster than the other only when there is enough evidence. We then modified Procedure \ref{alg:fa} by incorporating the test for significant difference using the three-way Compare function (Procedure \ref{alg:compare}), which is used to sort and rank algorithms into equivalence classes in Procedure \ref{alg:sort}. 
In Procedure \ref{alg:compare}, when the number of bootstrap iterations $M = 1$, the outcome that represents equivalence of algorithms ($\sim$) is no longer possible; and the outcome can just be either faster or  slower.
Table \ref{tab:1} shows the relative scores obtained by Procedure \ref{alg:f}  for $M=1$ ($threshold$ does not affect the outcome) and $M=30$ with different values of $threshold$.
% \p{Maybe it would be a good idea to repeat the meaning of
%  these scores}
%The experiments were also repeated by randomly switching between 20 to 24 threads to simulate system noise.
% The distributions of the algorithms in comparison with Algorithm 0  are shown in Figure \ref{fig:d}. 
 %Significant overlap can be observed between algorithms 0,1 and 2. 
 		% \begin{tabular}{c|ccc|ccc|}
 		% 	\cline{2-7}
 		% 	& \multicolumn{3}{|c|}{Without Noise} &  \multicolumn{3}{|c|}{With Noise} \\
 		% 	\cline{2-7}
 		% 	& Alg 0 & Alg 1 & Alg 2 & Alg 0 & Alg 1 & Alg 2 \\
 		% 	\hline
 		% 	\multicolumn{1}{ |c| }{M=1, thresh=N/A}  & 0.45 &0.11 &0.44 & 0.53 & 0.07 & 0.40 \\
 		% 	\hline
 		% 	\multicolumn{1}{ |c| }{M=30, thresh=0.50} & 0.53 &0.0 &0.47 & 0.76 & 0.0 & 0.24 \\
 		% 	\hline
 		% 	\multicolumn{1}{ |c| }{M=30, thresh=0.80} & 0.96 &0.73 &0.90 & 0.97 & 0.12 & 0.95\\
 		% 	\hline
 		% 	\multicolumn{1}{ |c| }{M=30, thresh=0.85}& 0.97 &0.89 &0.94 & 0.95 & 0.22 & 0.93 \\
 		% 	\hline
 		% 	\multicolumn{1}{ |c| }{M=30, thresh=0.90} & 0.99 &0.97 &0.99 & 0.95 & 0.66 & 0.90\\
 		% 	\hline
 		% 	\multicolumn{1}{ |c| }{M=30, thresh=0.95}& 1.0 &0.99 &0.99 & 0.97 & 0.88 & 0.97 \\
 		% 	\hline
 		% \end{tabular}
%\p{for all tables, please use tabbing and this next one as a template}
\begin{table}[h!]
  \begin{center}
    \renewcommand{\arraystretch}{1.2}
    \begin{tabular}{@{}r rrr c rrr@{}}
      \toprule
      & \multicolumn{3}{c}{Setting 1} & & \multicolumn{3}{c}{Setting 2} \\
      \cmidrule{2-4} \cmidrule{6-8}
                       & $\mathbf{a}_0$ & $\mathbf{a}_1$  & $\mathbf{a}_2$  && $\mathbf{a}_0$  & $\mathbf{a}_1$  & $\mathbf{a}_2$  \\
      \midrule
      {M=1, \hfill thr=N/A}
                       & 0.45  & 0.11  & 0.44  && 0.53  & 0.07  & 0.40  \\
      {M=30, thr=0.50} & 0.53  & 0.0   & 0.47  && 0.76  & 0.0   & 0.24  \\
      {M=30, thr=0.80} & 0.96  & 0.73  & 0.90  && 0.97  & 0.12  & 0.95  \\
      {M=30, thr=0.85} & 0.97  & 0.89  & 0.94  && 0.95  & 0.22  & 0.93  \\
      {M=30, thr=0.90} & 0.99  & 0.97  & 0.99  && 0.95  & 0.66  & 0.90  \\
      {M=30, thr=0.95} & 1.0   & 0.99  & 0.99  && 0.97  & 0.88  & 0.97  \\
      \bottomrule
    \end{tabular}
    \caption{Relative Scores; $Rep$ = 500, $K$ = 10}
    \label{tab:1}
  \end{center}
\end{table}
Relative scores close to 1 (or 0) implies that the algorithm consistently obtained (or did not obtain) rank 1 in all the $Rep$ repetitions of the sorting algorithm in Procedure \ref{alg:f}; therefore, as scores tend to 1, we consider the assignment of the algorithm to $\mathcal{F}$ to be more consistent or robust. When $M=1$,  $\mathbf{alg}_1$ received a low relative score, despite significantly overlapping with $\mathbf{alg}_0$ and $\mathbf{alg}_2$ in both settings. When $M=30$, bootstrapping is introduced; however, with $threshold=0.5$, the outcome ``$\sim$'' in the CompareAlg function is still not possible, but it can be seen that the score for $\mathbf{alg}_1$ is already 0 and can never be assigned to $\mathcal{F}$ (an indication of improved consistency).
%
%Therefore, based on the analysis from $N=50$ time measurements, $\mathbf{a}_1$ lagged slightly behind $\mathbf{a}_0$ and $\mathbf{a}_2$. However, 50 measurements are still only a snapshot of the true distribution, which is not available. Therefore, the test for significant
%difference is performed and the result of two comparisons is considered equivalent ($\sim$) if there is not enough evidence
%that one algorithm dominates the other. The tolerance level up to which two algorithms should be considered equivalent
%is controlled by adjusting the $threshold$ parameter in the Compare function. When $threshold=0.5$, the tolerance level
%is 0 and the outcome $\sim$ in the Compare function is still impossible. When the relative scores are computed with the
%setting  $M=30$ and $threshold=0.5$, $\mathbf{a}_0$ obtained the score 0 and was strictly not assigned to $\mathcal{F}$.
When the $threshold$ is increased from 0.5, the outcome ``$\sim$" in Procedure \ref{alg:sort} now becomes possible and as the $threshold$ tends to 1, the conditions for one algorithm to be ranked faster (or slower) than the other become stricter. As the $threshold$  increases, the relative score of $\mathbf{a}_1$ also
increases (see Table \ref{tab:1}), which means that it was assigned with rank 1 more frequently across different repetitions of the SortAlg function in Procedure \ref{alg:f}. Now all the algorithms $\mathbf{alg}_0, \mathbf{alg}_1, \mathbf{alg}_2$ with overlapping distributions of execution time gets relative score close to 1 as the $threshold$ increases, which implies that their assignment to $\mathcal{F}$ becomes more and more probable (robust). $\mathbf{alg}_3$ still obtains a relative score of 0; this is
because the difference in the distribution of execution time with the other algorithms is noticeable.

\begin{figure}[h!]
	\includegraphics[width=0.5\textwidth]{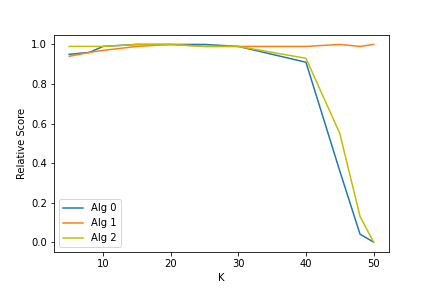}
	\caption{Relative Score vs $K$; $Rep$=500, $M$=30, $threshold$=0.9}
	\label{fig:k}     
\end{figure}

The parameter $threshold$ can be seen as something that controls the ``tolerance level" up to which two algorithms can be considered equivalent.  That is, a high $threshold$ implies that the tolerance to rank an algorithm better than the other is low and as a result, any two algorithms can result to be equivalent ($\sim$) unless there is enough evidence of one distribution dominating the other. The tolerance level is also impacted by the parameter $K$ in Procedure \ref{alg:compare}, which is the number of
measurements sampled from $N$ time measurements of an algorithm in each of the $M$ bootstrap iterations. Let $\mathbf{t}_i , \mathbf{t}_j \in \mathbb{R}^{N}$ be the time measurements of $\mathbf{alg}_i$ and $\mathbf{alg}_j$ respectively. In each of the $M$ iterations, let $(e_i,e_j)$ be the minimum execution time that is computed from
the samples $\mathbf{\tilde{t}_i} \subset \mathbf{t}_i$ and $\mathbf{\tilde{t}_j} \subset \mathbf{t}_j$ ($ \mathbf{\tilde{t}_i}, \mathbf{\tilde{t}_j} \in \mathbb{R}^{K}$). As
$K \to N$, the minimum of the samples $(\mathbf{\tilde{t}_i},\mathbf{\tilde{t}_j)}$ would tend to the minimum of the distributions $(\mathbf{t}_i, \mathbf{t}_j)$. As a consequence, when $K=N$, the evaluation of $e_i \le e_j$ in Procedure \ref{alg:compare} becomes deterministic, and ``$\sim$" would become an impossible outcome. For Setting 1 in Table \ref{tab:2}, $\mathbf{alg}_1$ recorded the lowest execution time; therefore, as $K \to N(=50)$, the relative score of $\mathbf{alg}_1$ approaches 1.0, while the scores of $\mathbf{alg}_0$ and $\mathbf{alg}_2$ approach 0 (see Figure \ref{fig:k}). 
Thus, higher values of $K$ invalidate the advantages of bootstrapping as the rankings start depending more and more exclusively on a single statistic. It is ideal to have $K$ randomly chosen from a set of values (say $K \in [5,10]$).

\subsection{Robustness to number of measurements}

We evaluate the assignments made to $\mathcal{F}$ as the number of time measurements $N$ of
each algorithm is reduced. All algorithms that obtain a relative score greater than 0 are assigned to $\mathcal{F}$. For
a given $N$, let $\mathcal{F}_{N}$ be the set of fastest algorithms that have been selected. As an example, let us
consider 100 mathematically equivalent algorithms %generated by the Linnea
framework for the Generalized Least Square problem~\cite{barthels2019linnea}; for $M=1$ (no bootstrapping), the set of fastest algorithms that are identified with $N=50$ is $\mathcal{F}_{50}:\{\mathbf{a}_0,\mathbf{a}_2\}$, and for $N=20$ is $\mathcal{F}_{20}:\{\mathbf{a}_0,\mathbf{a}_1,\mathbf{a}_2,\mathbf{a}_3,\mathbf{a}_4\}$; more algorithms are assigned to $\mathcal{F}_N$ when $N$ is reduced. When $N$ is decreased to 20, one finds less and less evidence of one
algorithm dominating over the others; as a consequence, more algorithms obtain rank 1 in at least one of the $Rep$
iterations of Procedure \ref{alg:fa} and are assigned to $\mathcal{F}_{20}$. We assume that $\mathcal{F}_{50}$ is
closer to the true solution than $\mathcal{F}_{N<50}$, as higher $N$ implies more evidence. 

%In Table~\ref{tab:3}, we report the average precision and recall (explained shortly) of
%$\mathcal{F}_{N<50} $ with respect to $\mathcal{F}_{50}$ for linear algebra expressions from 25 application examples
%(from \cite{barthels2019linnea}).
%We consider the solutions to be less robust when the precision is low. 
We use ``precision" and ``recall" metrics to evaluate consistency of the assignments made to $\mathcal{F}_N$. 
%In order to understand how we use these evaluation metrics, consider an instance for which rankings are obtained with $M=1$ (Procedure \ref{alg:fa})
%where
%$\mathcal{F}_{50}:\{\mathbf{a}_0,\mathbf{a}_2\}$ and
%$\mathcal{F}_{20}:\{\mathbf{a}_0,\mathbf{a}_1,\mathbf{a}_2,\mathbf{a}_3,\mathbf{a}_4\}$.
%When $N$ is decreased to 20, one finds less and less evidence of one
%algorithm dominating over the others; as a consequence, more algorithms manage to obtain rank 1 in at least one of the $T$
%iterations of Procedure \ref{alg:fa} and are assigned to $\mathcal{F}_{20}$.  We assume that $\mathcal{F}_{50}$ is the
%closer to the true solution than $\mathcal{F}_{N<50}$. 
When $\mathcal{F}_{20}$ is compared against $\mathcal{F}_{50}$,
the set $\mathcal{F}_{20}$ has  $\{\mathbf{a}_0,\mathbf{a}_2\}$ as true positives ($TP$),
$\{\mathbf{a}_1,\mathbf{a}_3,\mathbf{a}_4\}$  as false positives ($FP$), and there are no false negatives
($FN$). Therefore, the precision and recall for this example are 0.4 and 1.0, respectively.\footnote{precision =
  $TP/(TP+FP)$. recall = $TP/(TP+FN)$. Hence, $0\le$ precision $\le 1$, and $0\le$ recall $\le 1$.}
%We argue that for robustness, precision is a more important evaluation metric than recall; 
%When the precision is low but
%recall is high, although all the fastest algorithms are identified at a lower $N$ (resulting in a high recall score),
%$\mathcal{F}_{20}$ is cluttered with false positives. On the contrary, let us consider a case where precision is high and recall is low;  when $M=30$ and
%$threshold=0.5$, then $\mathcal{F}_{20} : \{\mathbf{a}_2\}$ contains one false negative $\{\mathbf{a}_0\}$ and no false positives; thus, the precision is 1.0 and
%the recall is 0.5. Although not all the fastest algorithms are identified (resulting in a lower recall score), there are
%no false positives and hence the solutions are precise.
%
% A low recall is still not ideal when consistent solutions are expected, but precision is a more important evaluation metric. 
\begin{table}[h!]
	\begin{center}
		\renewcommand{\arraystretch}{1.2}
		\begin{tabular}{@{}r rr c rr c rr c rr@{}}
			\toprule
			& \multicolumn{8}{c}{$M=30$} & & \multicolumn{2}{c}{$M=1$} \\
			\cmidrule{2-9} \cmidrule{11-12}
			& \multicolumn{2}{c}{thr=0.9} & & \multicolumn{2}{c}{thr=0.8} & & \multicolumn{2}{c}{thr=0.5} & & \multicolumn{2}{c}{thr=NA} \\
			\cmidrule{2-3} \cmidrule{5-6} \cmidrule{8-9} \cmidrule{11-12}
			{$N$} & \textbf{prc} & \textbf{rec} && \textbf{prc} & \textbf{rec} && \textbf{prc} & \textbf{rec} && \textbf{prc} & \textbf{rec} \\
			\midrule
			{40} & 0.97  & 0.94  && 0.86  & 0.90  && 0.71  & 0.67 && 0.32 & 0.99 \\
			{35} & 0.95  & 0.94  && 0.91  & 0.87  && 0.78  & 0.65 && 0.31 & 0.99 \\
			{30} & 0.93  & 0.86  && 0.88  & 0.87  && 0.68  & 0.58 && 0.34 & 0.99 \\
			{25} & 0.95  & 0.86  && 0.90  & 0.78  && 0.58  & 0.58 && 0.34 & 0.98 \\
			{20} & 0.97  & 0.80  && 0.93  & 0.75  && 0.50  & 0.38 && 0.36 & 0.95 \\
			{15} & 0.98  & 0.59  && 0.96  & 0.61  && 0.69  & 0.29 && 0.44 & 0.85 \\
			\bottomrule
		\end{tabular}
		\caption{Average Precision and Recall of $\mathcal{F}$; $Rep$=50, $K$=10}
		\label{tab:3}
	\end{center}
\end{table}
Table \ref{tab:3} shows the precision and recall values averaged for 25 different linear algebra expressions (taken from
\cite{barthels2019linnea}) arising in practical applications, and each expression consisting of at most 100 mathematically equivalent algorithms. It can be seen that the precision improves considerably when $M=30$ (with bootstrapping), even with zero tolerance (i.e., $threshold = 0.5$). Furthermore, both precision and recall improve as the $threshold$ increases. However, the recall decreases as $N$ is reduced. A lower recall score combined with high precision implies that, although not all the true fast algorithms are selected, the quality of the solutions identified is still preserved, as the number of false positives is less; in order to optimally select algorithms based on additional performance metric (such as energy), the set $\mathcal{F}$ should contain as little false positives as possible.

 However, we mentioned earlier that higher a $threshold$ implies that an algorithm will be ranked better than the other only when there is enough evidence of one distribution dominating the other; as $N$ decreases, the information about the actual distribution is also reduced, and therefore it is recommended to reduce the $threshold$ in proportion to $N$.

\section{Conclusion and Future Outlook}
\label{sec:con}

For a given mathematical expression, there typically exist many solution algorithms; although mathematically equivalent
to one another, those algorithms might exhibit significant differences in performance. We developed and evaluated a metric that quantifies the discrimination of an algorithm from the fastest algorithm. Our methodology is a measurement-based approach to identify not one, but a subset of algorithms that are equivalently fast\footnote{Code: https://github.com/as641651/Relative-Performance}. 
%We showed that the identifications are robust to system noise and the quality of solutions is preserved even when the number of measurements are decreased. 
This approach can be used when code optimization is done with respect to multiple factors such as execution time and energy or scalability, where identifying more than one fast algorithm becomes helpful~\cite{paiseSankaran}.

The typical development of scientific code involves a lot of trial and error in choosing the best implementation out
of several possible alternatives. In order to offer a high-level of abstraction,
%\p{I don't know what we're trying to say. Maybe ``In order to offer a high-level of abstraction'' ? ``In order to
%  facilitate programming''?}
languages such as Julia\cite{julia}, Matlab\cite{MatlabOTB}, Tensorflow\cite{tensorflow} etc., were developed to automatically decompose a mathematical expression into sequences of library calls. Compilers like Linnea~\cite{barthels2019linnea} were developed on top of such high-level languages to identify several alternative algorithms for the same mathematical expression. The selection of the best algorithm can be guided by models that predict relative performance. 

\paragraph*{\textbf{Relative Performance Modelling}} A natural extension to this work would be Relative Performance modelling, where we aim to automatically discriminate the algorithms without having to execute all the alternatives. To this end, our methodology can be used to develop machine learning models that are effective in doing what comes naturally to humans; that is, to learn through discrimination~\cite{alzubi2018machine}. If a human were assigned the task of selecting a fast algorithm among hundreds of alternatives, within a stipulated time-frame where he or she cannot execute and measure all those possible alternatives, a typical approach to solution would have been the following: First, sample a small set of algorithms and execute them; then, discriminate the faster algorithms from the slower algorithms by identifying the ``features'' (such as FLOPs, choice of BLAS calls, number of cores etc.) that resulted in an algorithm to execute faster or slower; finally, sample further algorithms for measurements, that have the appropriate features which could most likely result in those algorithm to execute faster. In order to develop a machine learning model for this task, our clustering methodology can be performed on a sample of algorithms and the relative scores can be mapped with the algorithm features to train a model iteratively, which can then be used to sample further algorithms that have high probabilities of being faster algorithms. We call this procedure as ``intelligent sampling'', for which a statistically sound relative performance metric is needed.

%\clearpage

\section*{Appendix}
\subsection{Algorithms for the Ordinary Least Square problem}
The algorithms are all mathematically equivalent; 
the FLOP count for \textbf{Algorithm 3} is twice as high as that for \textbf{Algorithm 0, 1, 2}.
\label{sec:appA}
\begin{algorithm}[H]
	\renewcommand{\thealgorithm}{}
	\floatname{algorithm}{Algorithm 0}
	\caption{ Blue }
	\label{alg:a0}
	\textbf{Expression: } $(X^TX)^{-1}X^{T}y \qquad X \in \mathbb{R}^{1000 \times 500} \quad y \in \mathbb{R}^{500}$ 
	\begin{algorithmic}[1] 
		\State $T_1 \leftarrow syrk(X^{T}X)$ \Comment{$T_1^{-1}X^{T}y$}
		\State $LL^{T} \leftarrow $ Cholesky($T_1$) \Comment{$L^{-T}L^{-1}X^{T}y$}
		\State $t_2 = X^{T}y$ \Comment{$L^{-T}L^{-1}t_2$}
		\State $t_2 = L^{-1}t_2$ \Comment{$L^{-T}t_2$}
		\State $z = L^{-T}t_2$
	\end{algorithmic}
\end{algorithm}

%In \textbf{Algorithm 1} and \textbf{Algorithm 2}, the order in which computations are evaluated is changed, but they have the same FLOP count  as \textbf{Algorithm 0}.
% Still, the order of computations can affect the execution time due of cache effects between sequence of calls\cite{peise2014cache}.
\begin{algorithm}[H]
	\renewcommand{\thealgorithm}{}
	\floatname{algorithm}{Algorithm 1}
	\caption{ Orange }
	\label{alg:a0}
	\textbf{Expression: } $(X^TX)^{-1}X^{T}y \qquad X \in \mathbb{R}^{1000 \times 500} \quad y \in \mathbb{R}^{500}$ 
	\begin{algorithmic}[1] 
		\State $t_1 = X^{T}y$ \Comment{$(X^{T}X)^{-1}t_1$}
		\State $T_2 \leftarrow syrk(X^{T}X)$ \Comment{$T_2^{-1}t_1$}
		\State $LL^{T} \leftarrow $ Cholesky($T_2$) \Comment{$L^{-T}L^{-1}t_1$}
		\State $t_1 = L^{-1}t_1$ \Comment{$L^{-T}t_1$}
		\State $z = L^{-T}t_1$
	\end{algorithmic}
\end{algorithm}

\begin{algorithm}[H]
	\renewcommand{\thealgorithm}{}
	\floatname{algorithm}{Algorithm 2}
	\caption{ Yellow }
	\label{alg:a0}
	\textbf{Expression: } $(X^TX)^{-1}X^{T}y \qquad X \in \mathbb{R}^{1000 \times 500} \quad y \in \mathbb{R}^{500}$ 
	\begin{algorithmic}[1] 
		\State $T_1 \leftarrow syrk(X^{T}X)$ \Comment{$T_1^{-1}X^{T}y$}
		\State $t_2 = X^{T}y$ \Comment{$T_1^{-1}t_2$}
		\State $LL^{T} \leftarrow $ Cholesky($T_1$) \Comment{$L^{-T}L^{-1}t_2$}
		\State $t_2 = L^{-1}t_2$ \Comment{$L^{-T}t_2$}
		\State $z = L^{-T}t_2$
	\end{algorithmic}
\end{algorithm}

\begin{algorithm}[H]
	\renewcommand{\thealgorithm}{}
	\floatname{algorithm}{Algorithm 3}
	\caption{ Red }
	\label{alg:a0}
	\textbf{Expression: } $(X^TX)^{-1}X^{T}y \qquad X \in \mathbb{R}^{1000 \times 500} \quad y \in \mathbb{R}^{500}$ 
	\begin{algorithmic}[1] 
		\State $T_1 \leftarrow gemm(X^{T}X)$ \Comment{$T_1^{-1}X^{T}y$}
		\State $LL^{T} \leftarrow $ Cholesky($T_1$) \Comment{$L^{-T}L^{-1}X^{T}y$}
		\State $t_2 = X^{T}y$ \Comment{$L^{-T}L^{-1}t_2$}
		\State $t_2 = L^{-1}t_2$ \Comment{$L^{-T}t_2$}
		\State $z = L^{-T}t_2$
	\end{algorithmic}

\end{algorithm}
%}

\bibliographystyle{IEEEtran}
\bibliography{pmbs2021}

\begin{thebibliography}{10}
\providecommand{\url}[1]{#1}
\csname url@samestyle\endcsname
\providecommand{\newblock}{\relax}
\providecommand{\bibinfo}[2]{#2}
\providecommand{\BIBentrySTDinterwordspacing}{\spaceskip=0pt\relax}
\providecommand{\BIBentryALTinterwordstretchfactor}{4}
\providecommand{\BIBentryALTinterwordspacing}{\spaceskip=\fontdimen2\font plus
\BIBentryALTinterwordstretchfactor\fontdimen3\font minus
  \fontdimen4\font\relax}
\providecommand{\BIBforeignlanguage}[2]{{%
\expandafter\ifx\csname l@#1\endcsname\relax
\typeout{** WARNING: IEEEtran.bst: No hyphenation pattern has been}%
\typeout{** loaded for the language `#1'. Using the pattern for}%
\typeout{** the default language instead.}%
\else
\language=\csname l@#1\endcsname
\fi
#2}}
\providecommand{\BIBdecl}{\relax}
\BIBdecl

\bibitem{peise2014cache}
E.~Peise and P.~Bientinesi, ``A study on the influence of caching: Sequences of
  dense linear algebra kernels,'' in \emph{International Conference on High
  Performance Computing for Computational Science}.\hskip 1em plus 0.5em minus
  0.4em\relax Springer, 2014, pp. 245--258.

\bibitem{hoefler2010characterizing}
T.~Hoefler, T.~Schneider, and A.~Lumsdaine, ``Characterizing the influence of
  system noise on large-scale applications by simulation,'' in \emph{SC'10:
  Proceedings of the 2010 ACM/IEEE International Conference for High
  Performance Computing, Networking, Storage and Analysis}.\hskip 1em plus
  0.5em minus 0.4em\relax IEEE, 2010, pp. 1--11.

\bibitem{peise2012performance}
E.~Peise and P.~Bientinesi, ``Performance modeling for dense linear algebra,''
  in \emph{2012 SC Companion: High Performance Computing, Networking Storage
  and Analysis}.\hskip 1em plus 0.5em minus 0.4em\relax IEEE, 2012, pp.
  406--416.

\bibitem{peise2019elaps}
------, ``The elaps framework: Experimental linear algebra performance
  studies,'' \emph{The International Journal of High Performance Computing
  Applications}, vol.~33, no.~2, pp. 353--365, 2019.

\bibitem{parasec}
\BIBentryALTinterwordspacing
C.~Bienia, S.~Kumar, J.~P. Singh, and K.~Li, ``The parsec benchmark suite:
  Characterization and architectural implications,'' in \emph{Proceedings of
  the 17th International Conference on Parallel Architectures and Compilation
  Techniques}, ser. PACT '08.\hskip 1em plus 0.5em minus 0.4em\relax New York,
  NY, USA: Association for Computing Machinery, 2008, p. 72–81. [Online].
  Available: \url{https://doi.org/10.1145/1454115.1454128}
\BIBentrySTDinterwordspacing

\bibitem{hoefler2015scientific}
T.~Hoefler and R.~Belli, ``Scientific benchmarking of parallel computing
  systems: twelve ways to tell the masses when reporting performance results,''
  in \emph{Proceedings of the international conference for high performance
  computing, networking, storage and analysis}, 2015, pp. 1--12.

\bibitem{tirer2018image}
T.~Tirer and R.~Giryes, ``Image restoration by iterative denoising and backward
  projections,'' \emph{IEEE Transactions on Image Processing}, vol.~28, no.~3,
  pp. 1220--1234, 2018.

\bibitem{barthels2019linnea}
H.~Barthels, C.~Psarras, and P.~Bientinesi, ``Linnea: Automatic generation of
  efficient linear algebra programs,'' \emph{arXiv preprint arXiv:1912.12924},
  2019.

\bibitem{psarras2019linear}
C.~Psarras, H.~Barthels, and P.~Bientinesi, ``The linear algebra mapping
  problem,'' \emph{arXiv preprint arXiv:1911.09421}, 2019.

\bibitem{julia}
J.~Bezanson, A.~Edelman, S.~Karpinski, and V.~B. Shah, ``Julia: A fresh
  approach to numerical computing,'' \emph{SIAM review}, vol.~59, no.~1, pp.
  65--98, 2017.

\bibitem{towardsEdgeComputing}
L.~{Lin}, X.~{Liao}, H.~{Jin}, and P.~{Li}, ``Computation offloading toward
  edge computing,'' \emph{Proceedings of the IEEE}, vol. 107, no.~8, pp.
  1584--1607, 2019.

\bibitem{connectedvehicles}
D.~Grewe, M.~Wagner, M.~Arumaithurai, I.~Psaras, and D.~Kutscher,
  ``Information-centric mobile edge computing for connected vehicle
  environments: Challenges and research directions,'' in \emph{Proceedings of
  the Workshop on Mobile Edge Communications}, 2017, pp. 7--12.

\bibitem{paiseSankaran}
\BIBentryALTinterwordspacing
A.~Sankaran and P.~Bientinesi, ``Performance comparison for scientific
  computations on the edge via relative performance,'' in \emph{{IEEE}
  International Parallel and Distributed Processing Symposium Workshops,
  {IPDPS} Workshops 2021, Portland, OR, USA, June 17-21, 2021}.\hskip 1em plus
  0.5em minus 0.4em\relax {IEEE}, 2021, pp. 887--895. [Online]. Available:
  \url{https://doi.org/10.1109/IPDPSW52791.2021.00132}
\BIBentrySTDinterwordspacing

\bibitem{bootstrap}
M.~R. Chernick, W.~Gonz{\'a}lez-Manteiga, R.~M. Crujeiras, and E.~B. Barrios,
  ``Bootstrap methods,'' 2011.

\bibitem{Balaprakash2018AutotuningIH}
P.~Balaprakash, J.~Dongarra, T.~Gamblin, M.~Hall, J.~Hollingsworth, B.~Norris,
  and R.~Vuduc, ``Autotuning in high-performance computing applications,''
  \emph{Proceedings of the IEEE}, vol. 106, pp. 2068--2083, 2018.

\bibitem{MatlabOTB}
``Matlab,'' 2019.

\bibitem{iakymchuk2012modeling}
R.~Iakymchuk and P.~Bientinesi, ``Modeling performance through memory-stalls,''
  \emph{ACM SIGMETRICS Performance Evaluation Review}, vol.~40, no.~2, pp.
  86--91, 2012.

\bibitem{iakymchuk2011execution}
------, ``Execution-less performance modeling,'' in \emph{Proceedings of the
  second international workshop on Performance modeling, benchmarking and
  simulation of high performance computing systems}, 2011, pp. 11--12.

\bibitem{agarwal2005impact}
S.~Agarwal, R.~Garg, and N.~K. Vishnoi, ``The impact of noise on the scaling of
  collectives: A theoretical approach,'' in \emph{International Conference on
  High-Performance Computing}.\hskip 1em plus 0.5em minus 0.4em\relax Springer,
  2005, pp. 280--289.

\bibitem{trackingPerfVariation}
J.~P.~S. Alcocer and A.~Bergel, ``Tracking down performance variation against
  source code evolution,'' \emph{ACM SIGPLAN Notices}, vol.~51, no.~2, pp.
  129--139, 2015.

\bibitem{robustbenchmarking}
J.~Chen and J.~Revels, ``Robust benchmarking in noisy environments,''
  \emph{arXiv preprint arXiv:1608.04295}, 2016.

\bibitem{statiscalperfCompare}
T.~Chen, Q.~Guo, O.~Temam, Y.~Wu, Y.~Bao, Z.~Xu, and Y.~Chen, ``Statistical
  performance comparisons of computers,'' \emph{IEEE Transactions on
  Computers}, vol.~64, no.~5, pp. 1442--1455, 2014.

\bibitem{hoefler2010loggopsim}
T.~Hoefler, T.~Schneider, and A.~Lumsdaine, ``Loggopsim: simulating large-scale
  applications in the loggops model,'' in \emph{Proceedings of the 19th ACM
  International Symposium on High Performance Distributed Computing}, 2010, pp.
  597--604.

\bibitem{davidWaitStates}
\BIBentryALTinterwordspacing
D.~B\"{o}hme, M.~Geimer, L.~Arnold, F.~Voigtlaender, and F.~Wolf, ``Identifying
  the root causes of wait states in large-scale parallel applications,''
  \emph{ACM Trans. Parallel Comput.}, vol.~3, no.~2, Jul. 2016. [Online].
  Available: \url{https://doi.org/10.1145/2934661}
\BIBentrySTDinterwordspacing

\bibitem{peise2014performance}
E.~Peise, D.~Fabregat-Traver, and P.~Bientinesi, ``On the performance
  prediction of blas-based tensor contractions,'' in \emph{International
  Workshop on Performance Modeling, Benchmarking and Simulation of High
  Performance Computer Systems}.\hskip 1em plus 0.5em minus 0.4em\relax
  Springer, 2014, pp. 193--212.

\bibitem{regressionScalability}
\BIBentryALTinterwordspacing
B.~J. Barnes, B.~Rountree, D.~K. Lowenthal, J.~Reeves, B.~de~Supinski, and
  M.~Schulz, ``A regression-based approach to scalability prediction,'' in
  \emph{Proceedings of the 22nd Annual International Conference on
  Supercomputing}, ser. ICS '08.\hskip 1em plus 0.5em minus 0.4em\relax New
  York, NY, USA: Association for Computing Machinery, 2008, p. 368–377.
  [Online]. Available: \url{https://doi.org/10.1145/1375527.1375580}
\BIBentrySTDinterwordspacing

\bibitem{barve2019fecbench}
Y.~D. Barve, S.~Shekhar, A.~Chhokra, S.~Khare, A.~Bhattacharjee, Z.~Kang,
  H.~Sun, and A.~Gokhale, ``Fecbench: A holistic interference-aware approach
  for application performance modeling,'' in \emph{2019 IEEE International
  Conference on Cloud Engineering (IC2E)}.\hskip 1em plus 0.5em minus
  0.4em\relax IEEE, 2019, pp. 211--221.

\bibitem{Jessup2016PerformanceBasedNS}
E.~Jessup, P.~Motter, B.~Norris, and K.~Sood, ``Performance-based numerical
  solver selection in the lighthouse framework,'' \emph{SIAM J. Sci. Comput.},
  vol.~38, 2016.

\bibitem{reiji}
R.~Suda, ``A bayesian method for online code selection: Toward efficient and
  robust methods of automatic tuning,'' \emph{Proc. Second International
  Workshop on Automatic Performance Tuning (IWAPT2007)}, 01 2007.

\bibitem{adaptiveOnline}
M.~Naghshnejad and M.~Singhal, ``Adaptive online runtime prediction to improve
  hpc applications latency in cloud,'' in \emph{2018 IEEE 11th International
  Conference on Cloud Computing (CLOUD)}, 2018, pp. 762--769.

\bibitem{comparingKrylov}
K.~Sood, B.~Norris, and E.~Jessup, ``Comparative performance modeling of
  parallel preconditioned krylov methods,'' in \emph{2017 IEEE 19th
  International Conference on High Performance Computing and Communications;
  IEEE 15th International Conference on Smart City; IEEE 3rd International
  Conference on Data Science and Systems (HPCC/SmartCity/DSS)}, 2017, pp.
  26--33.

\bibitem{kbest-kadioglu2011algorithm}
S.~Kadioglu, Y.~Malitsky, A.~Sabharwal, H.~Samulowitz, and M.~Sellmann,
  ``Algorithm selection and scheduling,'' in \emph{International Conference on
  Principles and Practice of Constraint Programming}.\hskip 1em plus 0.5em
  minus 0.4em\relax Springer, 2011, pp. 454--469.

\bibitem{peiseThesis}
\BIBentryALTinterwordspacing
E.~Peise, ``Performance modeling and prediction for dense linear algebra,''
  Dissertation, RWTH Aachen University, Aachen, 2017. [Online]. Available:
  \url{https://publications.rwth-aachen.de/record/721186/files/721186.pdf}
\BIBentrySTDinterwordspacing

\bibitem{bubblesort}
O.~Astrachan, ``Bubble sort: an archaeological algorithmic analysis,''
  \emph{ACM Sigcse Bulletin}, vol.~35, no.~1, pp. 1--5, 2003.

\bibitem{tensorflow}
M.~Abadi, P.~Barham, J.~Chen, Z.~Chen, A.~Davis, J.~Dean, M.~Devin,
  S.~Ghemawat, G.~Irving, M.~Isard \emph{et~al.}, ``Tensorflow: A system for
  large-scale machine learning,'' in \emph{12th $\{$USENIX$\}$ symposium on
  operating systems design and implementation ($\{$OSDI$\}$ 16)}, 2016, pp.
  265--283.

\bibitem{alzubi2018machine}
J.~Alzubi, A.~Nayyar, and A.~Kumar, ``Machine learning from theory to
  algorithms: an overview,'' in \emph{Journal of physics: conference series},
  vol. 1142, no.~1.\hskip 1em plus 0.5em minus 0.4em\relax IOP Publishing,
  2018, p. 012012.

\end{thebibliography}
\end{document}